\documentclass[reprint,amsmath,amssymb,aps,prc,floatfix,longbibliography]{revtex4-2}

\usepackage{amsmath}
\usepackage{graphicx}
\usepackage[caption=false]{subfig}

\usepackage{tikz-cd}
\usepackage{dcolumn}
\usepackage{bm}
\usepackage{multirow}
\usepackage{booktabs}
\usepackage{threeparttable}
\usepackage[colorlinks=true,linkcolor=blue,citecolor=blue,urlcolor=blue]{hyperref}

\begin{document}

\preprint{APS/123-QED}

\title{Background characterization in the PIKACHU Phase~1 search for two-neutrino double-beta decay of $^{160}$Gd}

\author{T. Omori$^{1}$}
\email{omori@hep.px.tsukuba.ac.jp}
\author{T. Iida$^{2}$}
\email{tiida@hep.px.tsukuba.ac.jp}

\author{A. Gando$^{3,4}$}
\author{N. Hinohara$^{5,2}$}
\author{K. Hosokawa$^{4}$}
\author{S. Ishidate$^{1}$}
\author{M. Ishigami$^{1}$}
\author{T. Kishimoto$^{6}$}
\author{K. Mizukoshi$^{4}$}
\author{T. Sako$^{1}$}
\author{S. Sugi$^{1}$}
\author{M. Yoshino$^{7,8}$}

\author{K. Fushimi$^{9}$}
\author{K. Kamada$^{7,8}$}
\author{Y. Shoji$^{8}$}
\author{Y. Suga$^{10}$}
\author{H. Suzuki$^{1}$}
\author{K. Takahashi$^{1}$}

\collaboration{PIKACHU Collaboration}

\affiliation{$^{1}$ Graduate School of Science and Technology, University of Tsukuba, Tsukuba, Ibaraki 305-8571, Japan}
\affiliation{$^{2}$ Institute of Pure and Applied Sciences, University of Tsukuba, Tsukuba, Ibaraki 305-8571, Japan}
\affiliation{$^{3}$ Department of Human Science, Obihiro University of Agriculture and Veterinary Medicine, Obihiro, Hokkaido 080-8555, Japan}
\affiliation{$^{4}$ Research Center for Neutrino Science, Tohoku University, Sendai, Miyagi 980-8578, Japan}
\affiliation{$^{5}$ Center for Computational Sciences, University of Tsukuba, Tsukuba, Ibaraki 305-8577, Japan}
\affiliation{$^{6}$ Research Center for Nuclear Physics, The University of Osaka, Ibaraki, Osaka 567-0047, Japan}
\affiliation{$^{7}$ New Industry Creation Hatchery Center, Tohoku University, Sendai, Miyagi 980-8579, Japan}
\affiliation{$^{8}$ C\&A Corporation, 1-16-23 Ichibancho, Aoba-ku, Sendai, Miyagi 980-0811, Japan}
\affiliation{$^{9}$ Division of Science and Technology, Tokushima University, Tokushima, Tokushima 770-8506, Japan}
\affiliation{$^{10}$ Mirion Technologies (Canberra) KK, Taito, Tokyo 111-0053, Japan}

\date{\today}
\begin{abstract}
We report a background characterization for the search for two-neutrino double-beta ($2\nu2\beta$) decay of $^{160}$Gd using Ce:Gd$_3$Al$_2$Ga$_3$O$_{12}$ scintillators in the PIKACHU experiment Phase~1 at the Kamioka underground laboratory. The experimental backgrounds were investigated through delayed-coincidence analysis and spectral fitting based on GEANT4 simulations. In particular, surface contamination of the crystal was identified through the characteristic distortion of the $^{214}\mathrm{Po}$ $\alpha$-ray spectrum, leading to an improved background model that reproduces the observed energy spectra. No significant excess attributable to $^{160}$Gd $2\nu2\beta$ decay was observed, yielding a lower limit of $T_{1/2}^{2\nu} \geq 2.32\times10^{19}$~yr at 90\% C.L. The established background model and analysis framework provide a foundation for future PIKACHU phases aiming to discover $^{160}$Gd $2\nu2\beta$ decay.
\end{abstract}

\maketitle

\section{Introduction}\label{sec:Introduction}
Neutrinoless double-beta ($0\nu2\beta$) decay is a process beyond the Standard Model that violates lepton number conservation. Its observation would establish the Majorana nature of neutrinos, demonstrating that neutrinos are their own antiparticles. The $0\nu2\beta$ decay half-life is related to the effective Majorana mass $m_{\beta\beta}$, providing an important probe of the absolute neutrino mass scale. However, the extraction of $m_{\beta\beta}$ from the measured half-life is limited by uncertainties in the nuclear matrix element (NME). Two-neutrino double-beta ($2\nu2\beta$) decay, in contrast, is an allowed process within the Standard Model and has been observed in several nuclei~\cite{Pritychenko2025}. Measurements of $2\nu2\beta$ half-lives provide important experimental benchmarks for NME calculations, helping to reduce their theoretical uncertainties and thereby improve the reliability of $m_{\beta\beta}$ extracted from $0\nu2\beta$ decay searches.

The PIKACHU (Pure Inorganic scintillator experiment in KAmioka for CHallenging Underground sciences) experiment aims to search for double-beta ($2\beta$) decay in $^{160}$Gd. This isotope has a relatively high natural abundance of 21.86(3)\%~\cite{Meija2016}, which is advantageous for rare-decay searches. However, its low $Q$-value of 1.7304(11)~MeV~\cite{Wang2021} makes the $2\beta$-decay search challenging, as radioactive backgrounds can contribute significantly in the signal energy region. Consequently, even $2\nu2\beta$ decay in $^{160}$Gd has not yet been observed. The best existing constraint on its half-life is $T_{1/2}^{2\nu} \geq 1.9\,(3.1)\times10^{19}$~yr at 90\% (68\%) C.L., obtained in a search using 103.6~g of $^{160}$Gd contained in a Ce-doped Gd$_2$SiO$_5$ (GSO) scintillator~\cite{Danevich2001}.

The PIKACHU experiment employs Ce-doped Gd$_3$Al$_2$Ga$_3$O$_{12}$ (GAGG) scintillators. These scintillators offer a high light yield exceeding 40000 photons/MeV~\cite{kamada2011}, approximately four times that of GSO~\cite{Takagi1983}, and their pulse shape discrimination (PSD) capability enables effective identification and rejection of $\alpha$-ray backgrounds. The GAGG scintillator is also well suited for large-scale searches, as large single crystals can be grown using the Czochralski method~\cite{Kochurikhin2020,Kamada2016}. These properties make GAGG a promising detector material for a low-background search for $2\beta$ decay in $^{160}$Gd.

Theoretical NME calculations predict $2\nu2\beta$ half-lives of $9.6\times10^{21}$~yr~\cite{Hirsch2002} and $7.4\times10^{20}$~yr~\cite{Hinohara2022} using the phase-space factor of Ref.~\cite{Kotila2012}. In addition, a semi-empirical approach yielded a predicted half-life of $3.4\times10^{20}$~yr~\cite{Pritychenko2023}. Reaching the theoretically predicted half-life region requires an improvement in experimental sensitivity of one to two orders of magnitude beyond the current limit. The PIKACHU experiment is therefore planned in two phases: Phase~1 aims to improve the experimental limit on the $^{160}$Gd $2\nu2\beta$ half-life, while Phase~2 aims to observe $2\nu2\beta$ decay or to constrain the NME by achieving sensitivity comparable to the theoretical predictions.

The present paper is organized as follows. The experimental setup is described in Sec.~\ref{sec:PIKACHUExperimentPhase1}, followed by the data processing and detector response in Secs.~\ref{sec:DataProcessing} and~\ref{sec:DetectorResponse}, respectively. The experimental background is characterized through delayed-coincidence analysis and spectral fitting in Sec.~\ref{sec:BackgroundStudy}. The resulting background model and a lower limit on the $^{160}$Gd $2\nu2\beta$ half-life are presented in Sec.~\ref{sec:Results}. Implications of the background characterization for future searches are discussed in Sec.~\ref{sec:Discussion}.

\section{PIKACHU Experiment Phase~1}\label{sec:PIKACHUExperimentPhase1}

\subsection{Experimental apparatus}\label{subsec:ExperimentalApparatus}
\begin{figure}[htbp]
\includegraphics[width=0.45\textwidth]{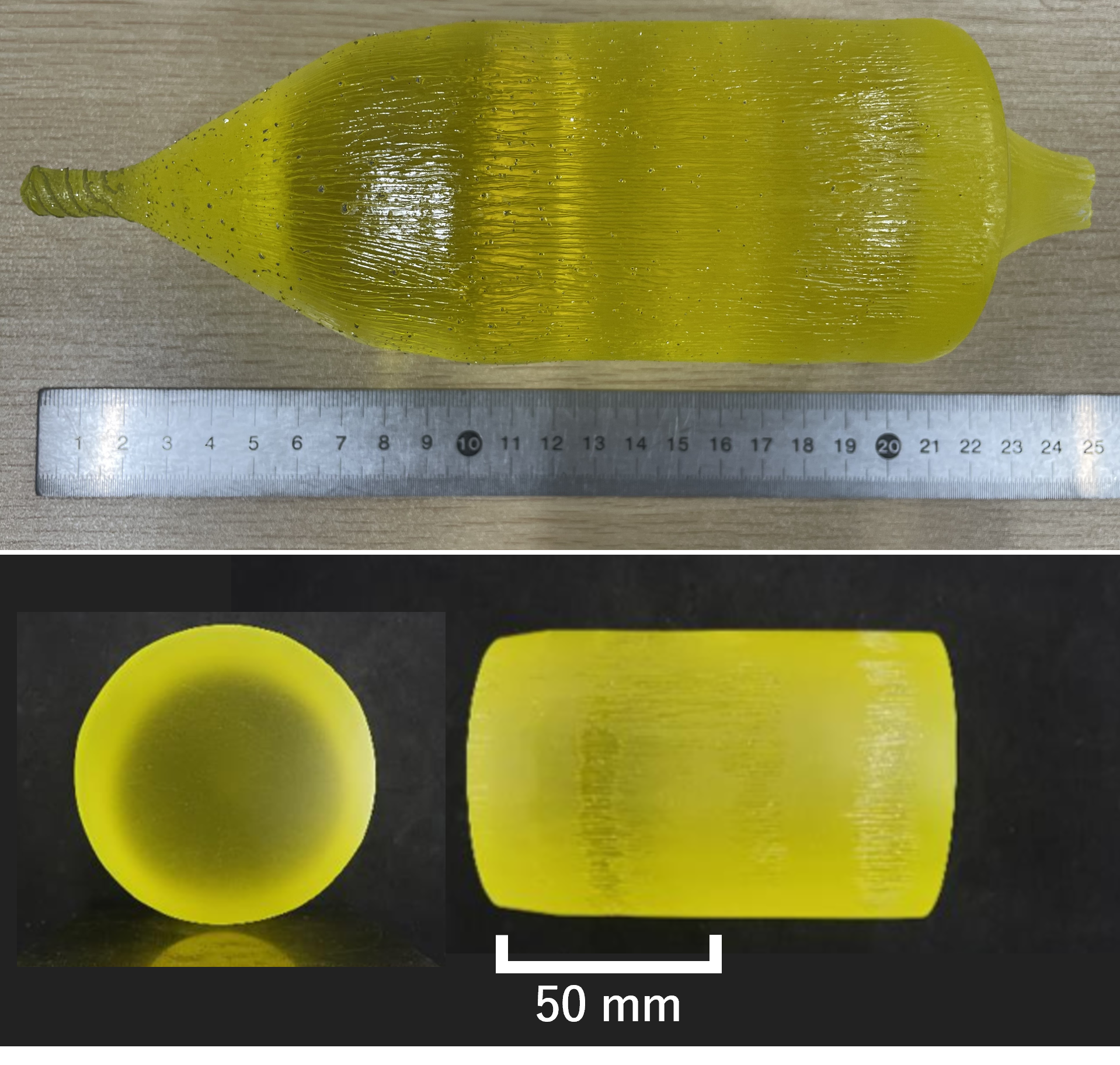}
\caption{
  \label{fig:Crystal} Photographs of Crystal~\#4 used in the PIKACHU experiment Phase~1. The top image shows the as-grown crystal, while the bottom images show the crystal after cutting (right) and mechanically polished cut surface (left).}
\end{figure}

In Phase~1 of the PIKACHU experiment, two GAGG scintillators, Crystal~\#1 and Crystal~\#4, were used. Crystal~\#1 had a diameter of 54~mm, a length of 52~mm, and a mass of 0.754~kg, whereas Crystal~\#4 had a diameter of 65~mm, a length of 100~mm, and a mass of 2.436~kg. Crystal~\#1 was the high-purity GAGG scintillator developed in Ref.~\cite{Omori2023}, while Crystal~\#4 was subsequently grown using the same purified raw materials as those used for Crystal~\#1. Photographs of Crystal~\#1 and Crystal~\#4 are shown in Fig.~2 in Ref.~\cite{Omori2023} and Fig.~\ref{fig:Crystal}, respectively. Based on the scintillator masses and the natural isotopic abundance of $^{160}$Gd, the masses of $^{160}$Gd contained in Crystal~\#1 and Crystal~\#4 were estimated to be 85~g and 276~g, respectively.

The scintillation light was collected by photomultiplier tubes (PMTs) through acrylic light guides (LGs). An R6231-04 PMT was used for Crystal~\#1, while an R6231-100 PMT was used for Crystal~\#4. Both PMTs, manufactured by Hamamatsu Photonics~\cite{Hamamatsu}, had a diameter of 50~mm and were otherwise identical in specification, except that the R6231-04 employed glass with a lower potassium content. Each LG had a truncated-cone shape, with diameters of 65~mm and 50~mm at the GAGG scintillator and PMT ends, respectively. The LGs were optically coupled to the scintillators and PMTs using optical grease. The surfaces of the scintillators and LGs were covered with PTFE reflectors and subsequently wrapped in black tape for light shielding.

\begin{figure*}[htbp]
\includegraphics[width=0.6\textwidth]{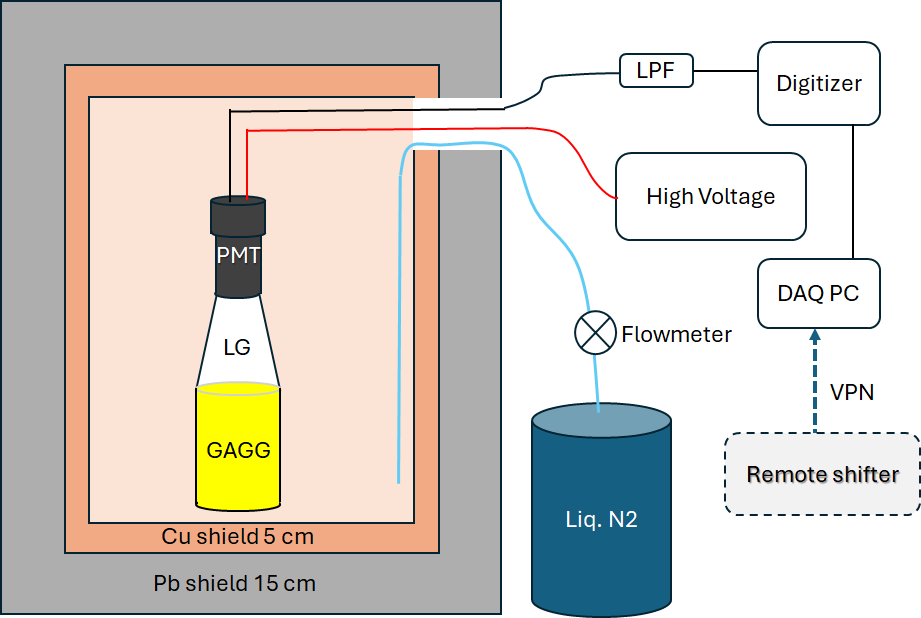}
\caption{\label{fig:Setup} Schematic overview of the detector setup, radiation shielding, and data acquisition system in Phase~1 of the PIKACHU experiment, showing the configuration with Crystal~\#4 installed.}
\end{figure*}

The detectors described above were installed in a radiation shield, shown in Fig.~\ref{fig:Setup}, at the Kamioka underground laboratory to reduce backgrounds from cosmic-ray muons and environmental radioactivity. The shield consisted of a 5~cm-thick copper layer surrounded by a 15~cm-thick lead layer.

The atmosphere in the Kamioka mine contains radon gas produced by the decay of radium in the surrounding rock~\cite{Pronost2018}. Radon and its decay products can contribute to the $\gamma$-ray background in the experiment. To mitigate this contribution, nitrogen gas was continuously introduced into the shield at a flow rate of 1~SLPM (standard liters per minute) to purge radon gas from the shield.

\subsection{Data acquisition}\label{subsec:DataAcquisition}
The PMT was operated at a bias voltage of $-1200$~V, and its output signal was passed through a low-pass filter (LPF) before digitization, as shown in Fig.~\ref{fig:Setup}. The LPF consisted of a 50~$\Omega$ resistor and a 100~pF capacitor, yielding a cutoff frequency of approximately 30~MHz. The LPF was introduced to suppress high-frequency spike noise in the PMT output and thereby enable stable data acquisition (DAQ) with low energy thresholds. The energy thresholds for $\alpha$ and $\beta/\gamma$ events were 220~keV and 130~keV, respectively, for Crystal~\#1, and 125~keV and 65~keV, respectively, for Crystal~\#4. The filtered signal was then digitized using a DT5720 digitizer manufactured by CAEN~\cite{CAEN}. The digitizer had a sampling rate of 250~MS/s, a resolution of 12 bits, and a dynamic range of 2~$\mathrm{V}_{\mathrm{pp}}$ (peak-to-peak voltage). In this work, the event window was set to 1024 samples, corresponding to a duration of 4096~ns.

The digitizer was controlled using CAENScope~\cite{CAENScope}, a data acquisition software package, on a DAQ PC located at the Kamioka underground laboratory. During data taking, data acquisition shifters remotely accessed the DAQ PC from their respective institutions to control and monitor the data acquisition. During Phase~1, data were first acquired with Crystal~\#4 for a total live time of 3102~h, followed by data acquisition with Crystal~\#1 for a total live time of 4173~h. During the data-taking period with Crystal~\#1, the PMT high voltage was switched off and subsequently turned on again. Accordingly, the Crystal~\#1 data were divided into two periods: Period~1, with a live time of 1950~h, and Period~2, with a live time of 2223~h. The spectral fits were performed independently for the two periods in Sec.~\ref{subsubsec:SpectralFittingProcedure}.
 
\section{Data Processing}\label{sec:DataProcessing}
This section describes the procedures used to process the waveform data acquired with the experimental setup described above. These procedures include energy reconstruction, energy calibration, and pulse shape discrimination (PSD) between $\alpha$ and $\beta/\gamma$ events.

\subsection{Energy reconstruction and calibration}\label{subsec:EnergyReconstructionAndCalibration}
\begin{figure}[htpb]
\includegraphics[width=0.45\textwidth]{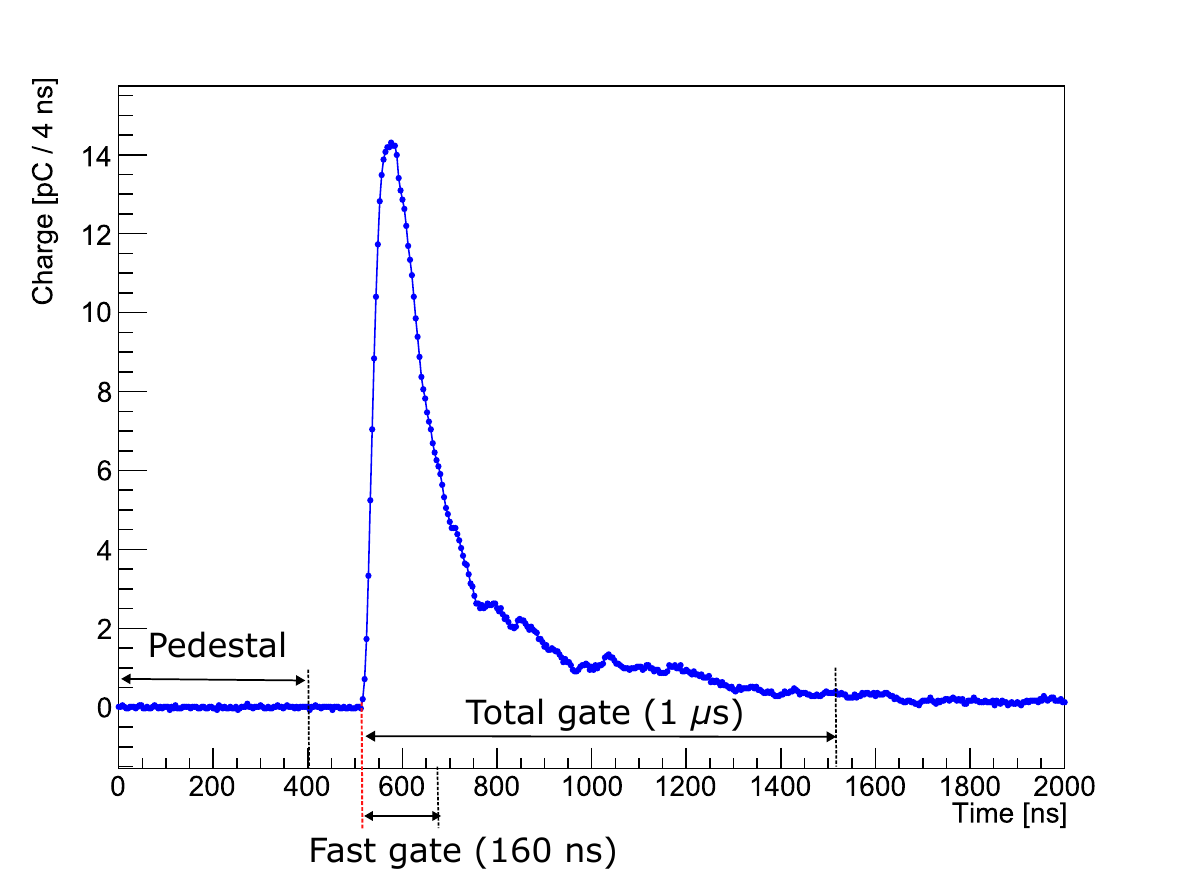}
\caption{\label{fig:Waveform}An example of a waveform recorded with Crystal~\#4. The vertical axis represents the charge per sampling interval (4~ns), calculated from the ADC counts after pedestal subtraction. The red dashed line indicates the rise time, determined using the pedestal level calculated from the waveform samples in the time interval shown in the figure. The ``Total gate'' indicates the integration interval used for energy reconstruction, while the integral over the ``Fast gate'' is used to obtain the fast component for pulse shape discrimination.}
\end{figure}

The energy of each event was reconstructed by integrating the corresponding GAGG waveform, an example of which is shown in Fig.~\ref{fig:Waveform}. First, the pedestal level and its standard deviation ($\sigma$) were calculated using the waveform samples in the time interval from 0 to 400~ns. The rise time, indicated by the red dashed line in Fig.~\ref{fig:Waveform}, was defined as the first time at which the waveform amplitude exceeded the pedestal level by more than $10\sigma$ for five consecutive samples. The waveform was then integrated over a 1~$\mu$s window starting at the rise time. This integration interval corresponds to the ``Total gate'' shown in Fig.~\ref{fig:Waveform}.

The energy scale for $\alpha$ events was calibrated using the 1274~keV photopeak in the $^{22}$Na $\gamma$-ray spectrum, measured immediately after each data-taking period. This calibration was adopted because the nonproportional responses of both scintillators were evaluated relative to their responses to 1274~keV $\gamma$ rays~\cite{Omori2026_2}. The resulting nonproportional responses were taken into account when determining the expected energies of the de-excitation $\gamma$ rays emitted following $\alpha$ decays in the background modeling (see Sec.~\ref{subsubsec:MonteCarloSimulation}). In contrast, the energy scale for $\beta/\gamma$ events was calibrated in situ using the 1460~keV and 2614~keV $\gamma$-ray peaks from $^{40}$K and $^{208}$Tl, respectively, observed in each dataset.

\subsection{Pulse shape discrimination}\label{subsec:PulseShapeDiscrimination}
To distinguish $\alpha$ and $\beta/\gamma$ events, a pulse-shape parameter, hereafter referred to as the Indicator, was defined as
\begin{equation}
    \mathrm{Indicator} = \frac{Q_{\mathrm{fast}}}{Q_{\mathrm{total}}},
\end{equation}
where $Q_{\mathrm{fast}}$ is the integrated waveform area within 160~ns from the rise time, corresponding to the ``Fast gate'' in Fig.~\ref{fig:Waveform}, and $Q_{\mathrm{total}}$ is the total waveform area, corresponding to the reconstructed energy defined in Sec.~\ref{subsec:EnergyReconstructionAndCalibration}.

\begin{figure*}[htbp]
    \centering
    \subfloat[Crystal~\#1\label{fig:PSDCrystal1}]{%
        \includegraphics[width=0.49\textwidth]{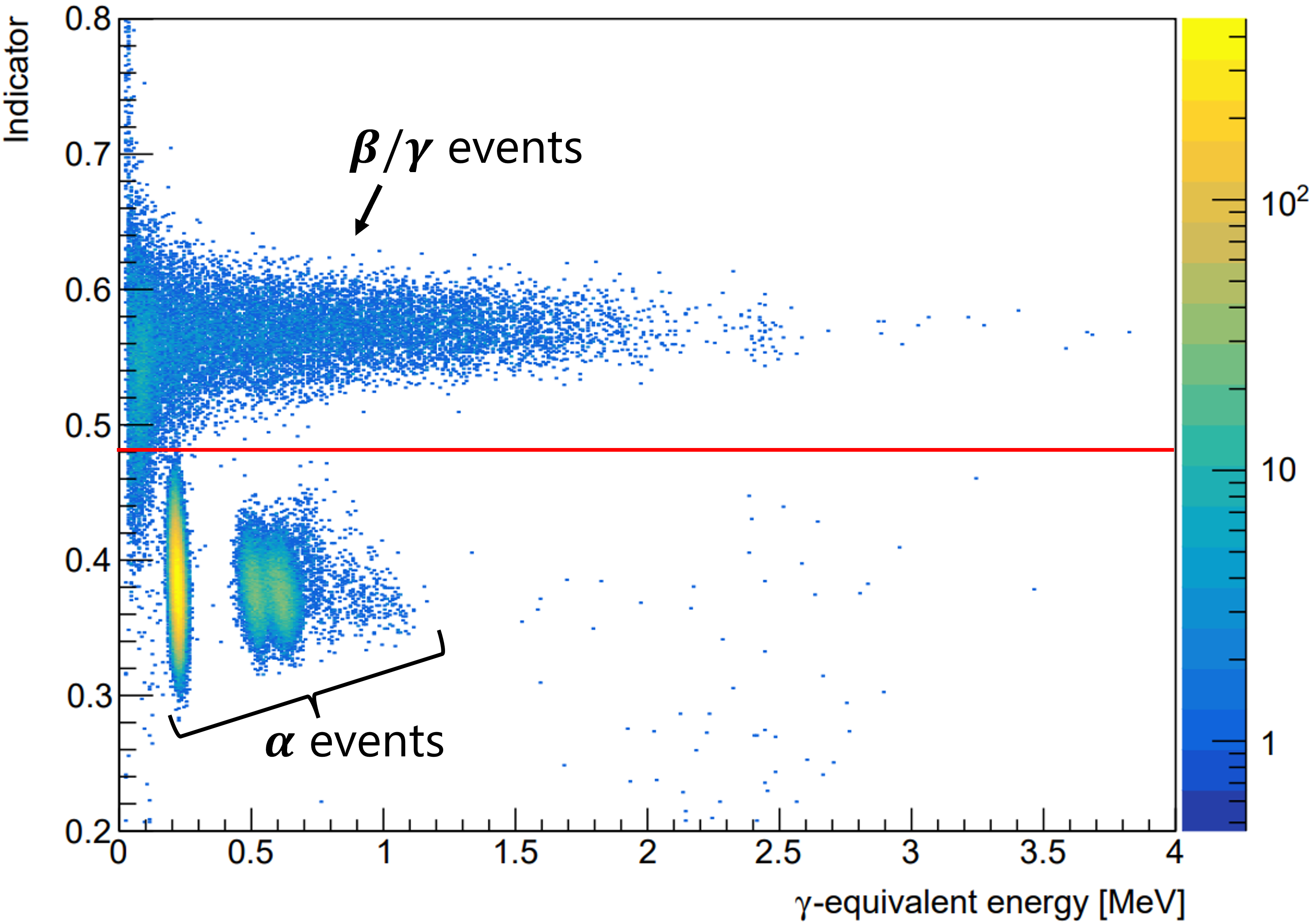}%
    }\hfill
    \subfloat[Crystal~\#4\label{fig:PSDCrystal4}]{%
        \includegraphics[width=0.49\textwidth]{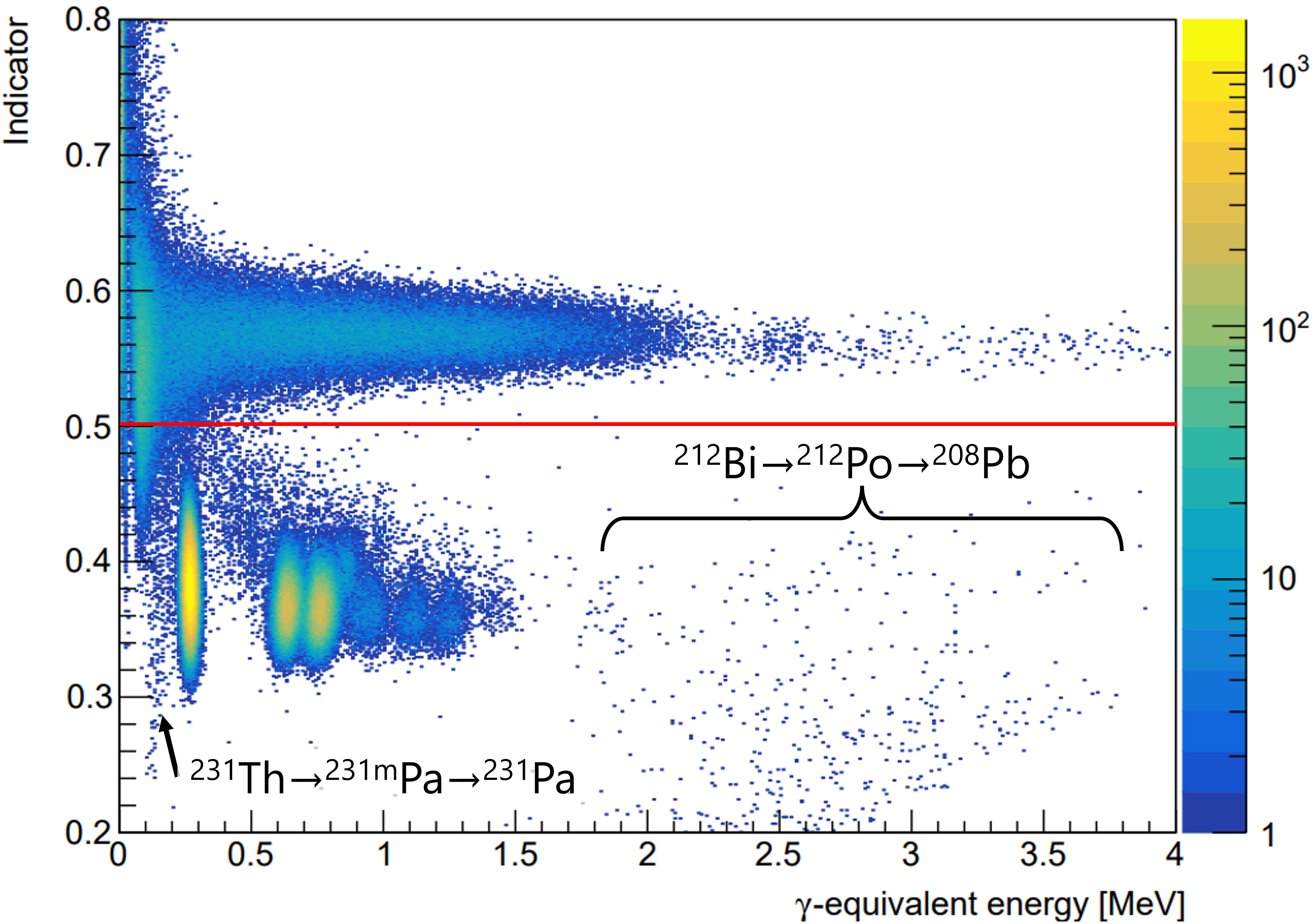}%
    }

    \caption{
      Two-dimensional distributions of the Indicator versus the $\gamma$-equivalent energy for one-day datasets. The Indicator is shown on the $y$-axis, and the $\gamma$-equivalent energy is shown on the $x$-axis. The red horizontal lines indicate the Indicator thresholds used to distinguish $\alpha$ and $\beta/\gamma$ events in the subsequent analysis. The sequential decays $^{231}\mathrm{Th}\rightarrow{}^{231\mathrm{m}}\mathrm{Pa}\rightarrow{}^{231}\mathrm{Pa}$ and $^{212}\mathrm{Bi}\rightarrow{}^{212}\mathrm{Po}\rightarrow{}^{208}\mathrm{Pb}$ appear as pileup events because the successive decays can occur within a single event window.}
    \label{fig:PSD}
\end{figure*}

Figure~\ref{fig:PSD} shows the two-dimensional distributions of the Indicator versus the $\gamma$-equivalent energy for one-day datasets. In this work, events with Indicator values below 0.48 and 0.50 were selected as $\alpha$ events for Crystal~\#1 and Crystal~\#4, respectively, while events with Indicator values above these thresholds were selected as $\beta/\gamma$ events.

The dominant $\alpha$-event populations observed at 0.2--0.3~MeV originate from $^{152}$Gd and $^{147}$Sm. The $\alpha$ events from the uranium and thorium (U/Th) decay chains are distributed mainly in the 0.5--1.5~MeV region. The events observed at lower energies, adjacent to the $^{152}$Gd and $^{147}$Sm populations, originate from the $\beta$ decay of $^{231}$Th to an excited state of $^{231}$Pa, followed by its de-excitation. The excited state, denoted as $^{231\mathrm{m}}$Pa, has a half-life of 45.1~ns.

Furthermore, events distributed over the 1.5--4.0~MeV in the $\alpha$-event region originate from the sequential decays of $^{212}$Bi and $^{212}$Po in the $^{232}$Th decay chain: $^{212}\mathrm{Bi}\left(T_{1/2}=60.55~\mathrm{min}\right)\rightarrow{}^{212}\mathrm{Po}\left(T_{1/2}=299~\mathrm{ns}\right)\rightarrow{}^{208}\mathrm{Pb}$. Since the half-life of $^{212}$Po is shorter than the event window, the two successive decays are recorded as a single pileup event. These pileup events are clearly visible in Crystal~\#4 in Fig.~\ref{fig:PSD}. Although the corresponding population is not clearly visible in the Crystal~\#1 distribution because of the limited statistics, the isotopes responsible for these sequential decays are also present in Crystal~\#1.

In Crystal~\#4, some overlap between $\alpha$ and $\beta/\gamma$ events is observed below 400~keV, and the pileup events described above are included in the $\alpha$-event population. However, these contributions have a negligible impact on the subsequent analyses because they are largely excluded from the energy regions used in those analyses.

\begin{figure}[htbp]
\includegraphics[width=0.45\textwidth]{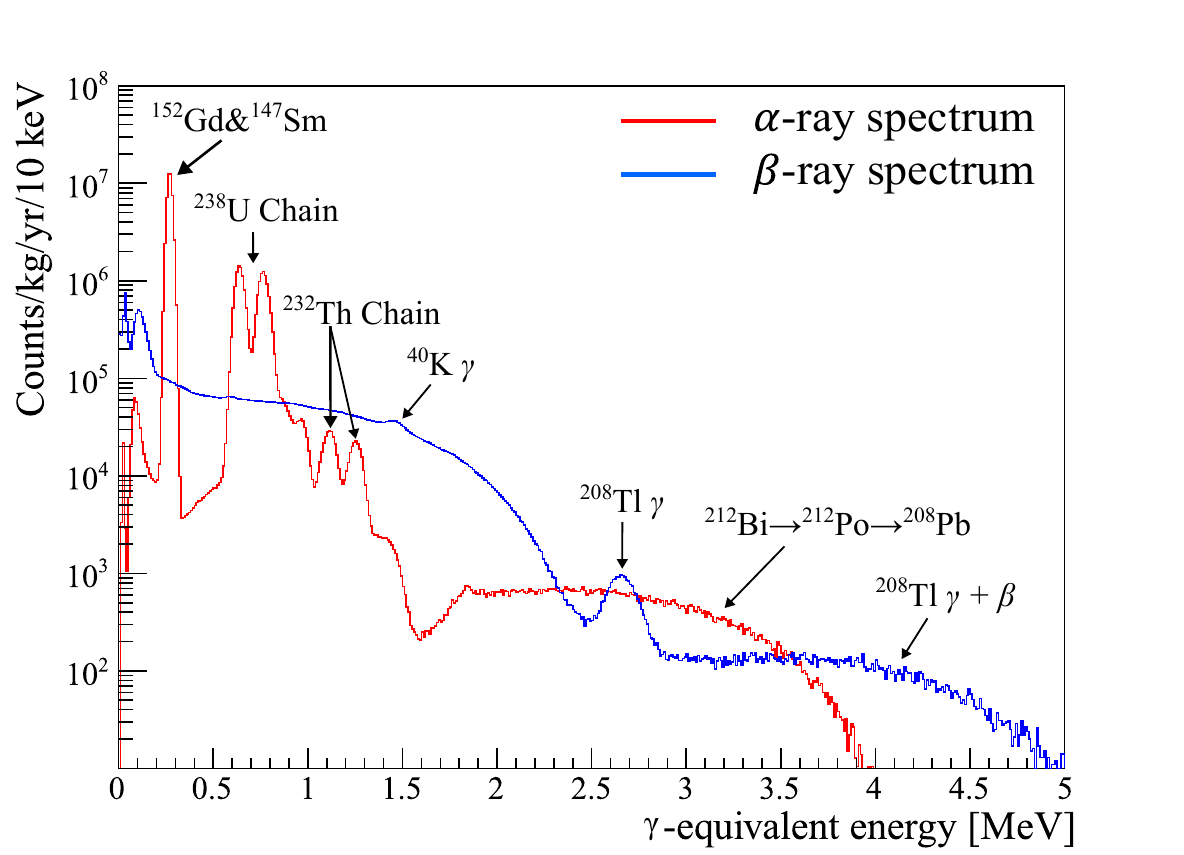}
\caption{\label{fig:Spectrum}Energy spectra obtained from the 3102~h dataset acquired with Crystal~\#4. The red and blue lines represent the $\alpha$- and $\beta/\gamma$-ray spectra, respectively, separated by PSD. The source isotopes responsible for the prominent peaks are indicated in the figure.}
\end{figure}

By applying the energy calibration and PSD described above to each dataset, the $\alpha$- and $\beta/\gamma$-ray energy spectra were obtained. Figure~\ref{fig:Spectrum} shows the energy spectra for the 3102~h dataset acquired with Crystal~\#4 as a representative example. The red and blue lines represent the $\alpha$- and $\beta/\gamma$-ray spectra, respectively. The event rates are normalized by the live time, crystal mass, and energy bin width.

\section{Detector Response}\label{sec:DetectorResponse}
An accurate understanding of the detector response is essential for constructing reliable background and signal models. In particular, the observed spectral shapes are affected by the detector's energy resolution, quenching, and nonproportionality. This section describes the evaluation of these detector-response characteristics, which are subsequently incorporated into the background and signal models.

\subsection{Energy resolution}\label{subsec:EnergyResolution}
The energy resolution of the detector for $\alpha$ rays was evaluated using the unresolved peak from $^{152}$Gd and $^{147}$Sm decays and the $^{238}$U $\alpha$-decay peaks observed in each $\alpha$-ray spectrum, as well as the $^{220}$Rn and $^{216}$Po $\alpha$-decay peaks identified by the delayed-coincidence (DC) analysis described in Sec.~\ref{subsec:DelayedCoincidenceAnalysis}. In contrast, the energy resolution for $\beta/\gamma$ events was evaluated using the photopeaks of $\gamma$ rays emitted by $^{22}$Na, $^{40}$K, $^{60}$Co, $^{133}$Ba, $^{137}$Cs, $^{208}$Tl, and $^{241}$Am sources. The $^{40}$K source consisted of potassium chloride powder contained in a sample container, while the $^{208}$Tl source was a 2\% thoriated tungsten electrode used for tungsten inert gas welding. Data from these $\gamma$-ray sources were acquired immediately after each data-taking period.

\begin{figure*}[htbp]
    \centering
    \subfloat[Resolution for $\alpha$ rays\label{fig:AlphaResolution}]{%
        \includegraphics[width=0.49\textwidth]{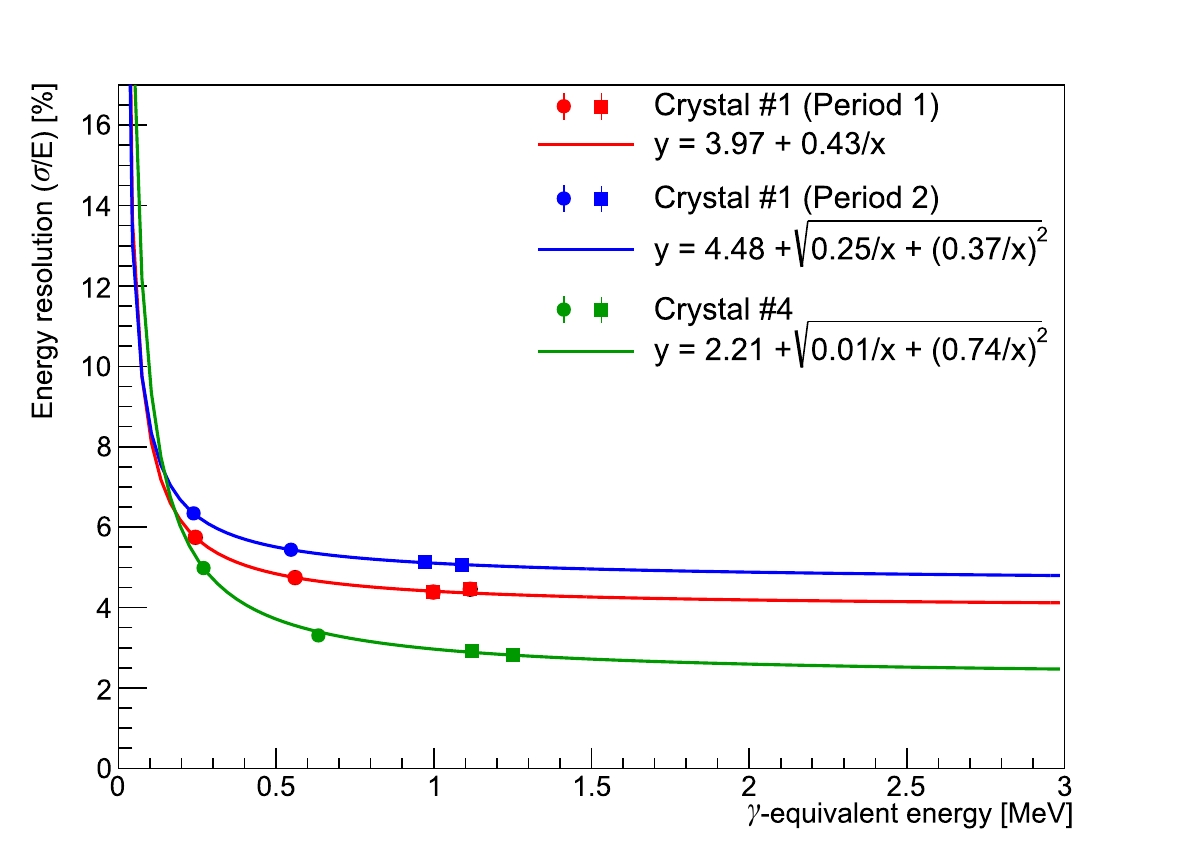}%
    }\hfill
    \subfloat[Resolution for $\gamma$ rays\label{fig:GammaResolution}]{%
        \includegraphics[width=0.49\textwidth]{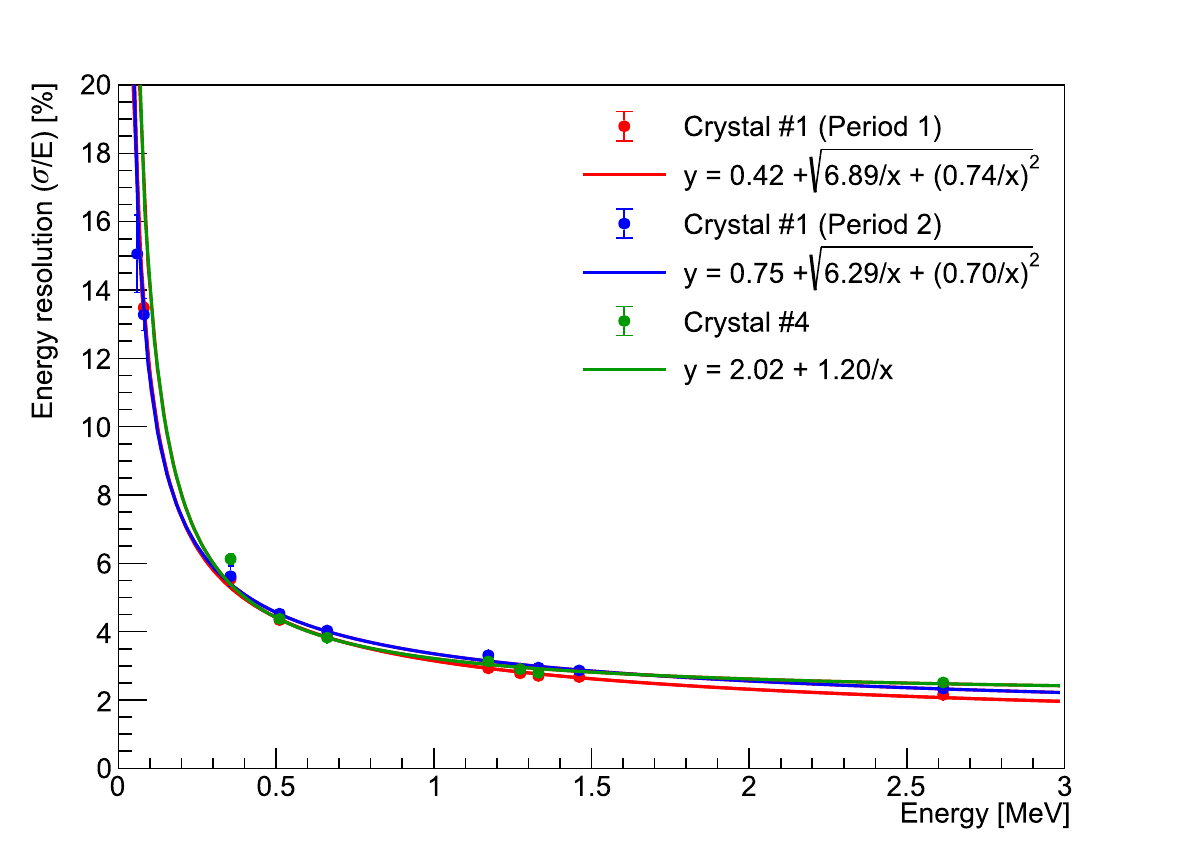}%
    }

    \caption{
      Energy resolution as a function of energy for (a) $\alpha$ rays and (b) $\gamma$ rays. The energy resolution is defined as $\sigma/E$. In panel (a), the circle and square markers represent the resolutions evaluated from peaks in the $\alpha$-ray spectra and from the $^{220}$Rn and $^{216}$Po peaks identified by the delayed-coincidence analysis, respectively. Different colored markers distinguish the three datasets: Crystal~\#1 Period~1, Crystal~\#1 Period~2, and Crystal~\#4. The colored curves represent the parameterizations obtained using Eq.~\eqref{eq:EnergyResolution}.
      }
    \label{fig:Resolution}
\end{figure*}

Figures~\ref{fig:AlphaResolution} and~\ref{fig:GammaResolution} show the energy resolutions for $\alpha$ and $\gamma$ rays, respectively, as a function of the energy. The energy resolution is defined as $\sigma/E$ and is expressed as a percentage. In Fig.~\ref{fig:AlphaResolution}, the circle and square markers represent the resolutions evaluated from peaks in the $\alpha$-ray spectra and from the $^{220}$Rn and $^{216}$Po peaks identified by the DC analysis, respectively. The energy resolutions of Crystal~\#1 differ between Periods~1 and 2. Although the origin of this difference has not been identified, possible causes include changes in the optical coupling conditions between the detector components.

The energy dependences of the resolutions for $\alpha$ and $\gamma$ rays were parameterized using the following function:
\begin{equation}
y = p_0+\sqrt{\frac{p_1}{x}+\left(\frac{p_2}{x}\right)^2},
\label{eq:EnergyResolution}
\end{equation}
where $y$ denotes the energy resolution, $\sigma/E$, expressed as a percentage, and $x$ represents the energy. The parameters $p_0$, $p_1$, and $p_2$ are free parameters. The resulting parameterizations are shown by the colored curves in Fig.~\ref{fig:Resolution}.

\subsection{Quenching response for $\alpha$ rays}\label{subsec:QuenchingResponseForAlphaRays}
Quenching refers to the reduced scintillation response of $\alpha$ rays compared with that of $\gamma$ rays of the same energy. In GAGG scintillators, the $\alpha/\gamma$ ratio, defined as the ratio of the $\gamma$-equivalent energy to the original $\alpha$-ray energy, is known to depend linearly on the original $\alpha$-ray energy~\cite{Furuno2021}. Accordingly, the $\gamma$-equivalent energy of $\alpha$ rays was parameterized as a quadratic function of the original $\alpha$-ray energy in this work.

\begin{figure}[htbp]
\includegraphics[width=0.45\textwidth]{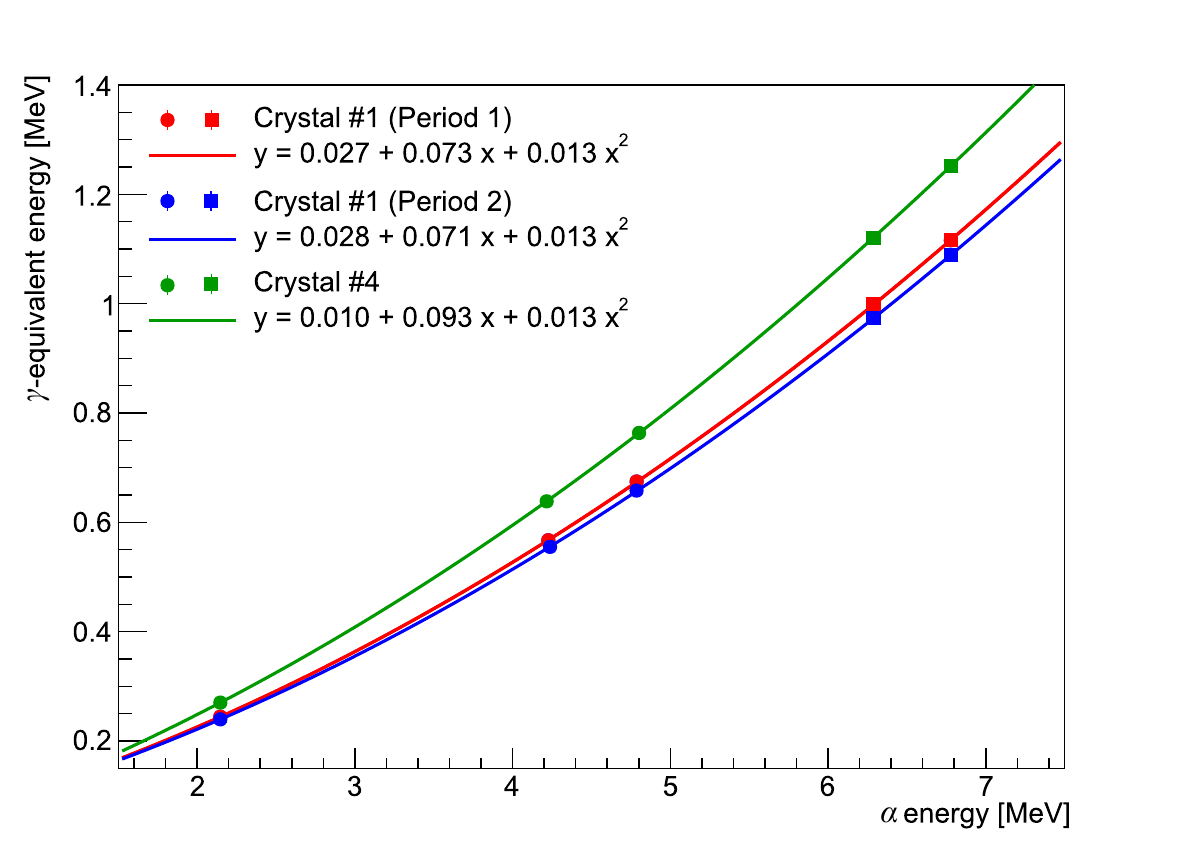}
\caption{\label{fig:Quenching} Quenching response of the GAGG scintillators for $\alpha$ rays. The observed $\gamma$-equivalent energy is plotted as a function of the original $\alpha$-ray energy. The circle and square markers represent the peaks identified in the $\alpha$-ray spectra and by the delayed-coincidence analysis, respectively. Different colors distinguish the three datasets: Crystal~\#1 Period~1, Crystal~\#1 Period~2, and Crystal~\#4. The colored curves represent the quadratic fits to the data for each dataset.}
\end{figure}

Figure~\ref{fig:Quenching} shows the observed $\gamma$-equivalent energies of $\alpha$-ray peaks as a function of their original energies. The circle markers represent the $^{152}\mathrm{Gd}$/$^{147}\mathrm{Sm}$, $^{238}$U, and $^{234}$U/$^{230}$Th peaks observed in the $\alpha$-ray spectra, while the square markers represent the $^{220}$Rn and $^{216}$Po peaks identified by the DC analysis. The colored curves show the quadratic fits to the data for each dataset.

\subsection{Nonproportional response for $\beta$ and $\gamma$ rays}\label{subsec:NonproportionalResponseForBetaAndGammaRays}
At low energies, GAGG scintillators have been reported to exhibit a nonproportional response in their scintillation light yield~\cite{Sibczynski2018,Kaewkhao2016}. We previously evaluated the nonproportional response to $\beta$ and $\gamma$ rays of Crystal~\#4, using the Compton coincidence technique and several $\gamma$-ray sources, respectively, over an energy range of 50--2614~keV~\cite{Omori2026_2}. In Ref.~\cite{Omori2026_2}, the nonproportional response was characterized by nPR, defined as the ratio of the observed energy to the deposited energy ($E_{\mathrm{dep}}$). The measured nPR values for $\beta$ and $\gamma$ rays were parameterized as double-exponential functions of $E_{\mathrm{dep}}$. For this work, we also evaluated the nonproportional response of Crystal~\#1 using the same methods. The nPR values for $\beta$ and $\gamma$ rays as functions of $E_{\mathrm{dep}}$, together with the corresponding fitting results for Crystal~\#4, are shown in Fig.~6 of Ref.~\cite{Omori2026_2}.

\section{Background Study}\label{sec:BackgroundStudy}
To accurately characterize the experimental backgrounds in Phase~1, we performed a DC analysis and Monte Carlo simulations. The DC analysis identified specific radioactive decay chains through their characteristic time correlations and provided independent measurements of their activities, while the Monte Carlo simulations quantified the contributions of these radioactive contaminants to the observed energy spectra.

\subsection{Delayed-coincidence analysis}\label{subsec:DelayedCoincidenceAnalysis}
In the U/Th decay chains, several successive decays occur on millisecond or sub-millisecond timescales. Such decay sequences can be identified by searching for successive events using the timestamps recorded by the digitizer, which has a time resolution of 8~ns. In this section, three decay sequences identified by the DC analysis are described.

\subsubsection{$^{220}\mathrm{Rn}\rightarrow^{216}\mathrm{Po}\rightarrow^{212}\mathrm{Pb}$ in the $^{232}$Th decay chain}\label{subsubsec:232ThDecayChain}

The relevant part of the $^{232}$Th decay chain is

\begin{equation}
\begin{aligned}
&{}^{220}\mathrm{Rn}\left(T_{1/2}=55.6(1)~\mathrm{s}\right)\rightarrow \\
&\qquad {}^{216}\mathrm{Po}\left(T_{1/2}=144.0(6)~\mathrm{ms}\right)\rightarrow
{}^{212}\mathrm{Pb}.
\end{aligned}
\end{equation}

The $^{220}$Rn and $^{216}$Po decays emit $\alpha$ particles with energies of 6.288~MeV and 6.778~MeV, respectively. The $^{220}$Rn decay was selected as the trigger events by requiring an $\alpha$-ray energy within $E_{\mathrm{vis}}(6.288)\pm3\sigma_{\alpha}$, where $E_{\mathrm{vis}}(6.288)$ is the $\gamma$-equivalent energy corresponding to a 6.288-MeV $\alpha$ particle, calculated using the quenching function shown in Fig.~\ref{fig:Quenching}. Here, $\sigma_{\alpha}$ denotes the energy resolution for $\alpha$ rays shown in Fig.~\ref{fig:Resolution}. For each trigger event, subsequent $\alpha$ events with energies within $E_{\mathrm{vis}}(6.778)\pm3\sigma_{\alpha}$ were searched for within 2000~ms after the trigger. The time difference between the trigger and each subsequent event was then calculated.

Figure~\ref{fig:Tdiff232Th} shows the resulting time-difference distribution for Crystal~\#4. The distribution was fitted with the sum of an exponential function representing the $^{216}$Po decay and a constant function representing random coincidences. The red curve shows the total fit, while the solid and dashed black curves represent the exponential and constant components, respectively. The half-life obtained from the fit was $143.8\pm0.5$~ms, in good agreement with the reference value. The total number of $^{216}$Po decays was obtained by integrating the fitted exponential component from $t=0$ to $\infty$; therefore, no additional correction for the finite time window was required. After correcting for the combined energy-selection efficiency of $\eta=0.994$ for the trigger and subsequent events, the $^{216}$Po activity was determined to be $5.37\pm0.03$~mBq/kg. The half-lives and activities obtained by the same analysis for Crystal~\#1 data are summarized in Table~\ref{tab:DCSummary}.

The $\alpha$-ray energy spectra of the $^{220}$Rn and $^{216}$Po events selected by this analysis for Crystal~\#4 are shown by the black and gray curves, respectively, in Fig.~\ref{fig:DCSpectrum}. Similar peaks were also observed in Crystal~\#1, but are omitted from the figure for clarity.

\subsubsection{$^{219}\mathrm{Rn}\rightarrow^{215}\mathrm{Po}\rightarrow^{211}\mathrm{Pb}$ in the $^{235}$U decay chain}
A similar sequence occurs in the $^{235}$U decay chain:

\begin{equation}
\begin{aligned}
&{}^{219}\mathrm{Rn}\left(T_{1/2}=3.96(1)~\mathrm{s}\right)\rightarrow \\
&\qquad {}^{215}\mathrm{Po}\left(T_{1/2}=1.781(5)~\mathrm{ms}\right)\rightarrow
{}^{211}\mathrm{Pb}.
\end{aligned}
\end{equation}

The $^{219}$Rn and $^{215}$Po decays emit $\alpha$ particles with energies of 6.819~MeV (79.4\% intensity) and 7.386~MeV, respectively. Following the procedure described in Sec.~\ref{subsubsec:232ThDecayChain}, the $^{219}$Rn decay was selected as the trigger by requiring an $\alpha$-ray energy within $E_{\mathrm{vis}}(6.819)\pm1\sigma_{\alpha}$. Subsequent $\alpha$ events with energies within $E_{\mathrm{vis}}(7.386)\pm3\sigma_{\alpha}$ were searched for within 0--15~ms after the trigger.

The resulting time-difference distribution for Crystal~\#4 is shown in Fig.~\ref{fig:Tdiff235U}. It was fitted with an exponential function and a constant component to account for random coincidences. The half-life obtained from the fit was $1.55\pm0.46$~ms, which is consistent with the reference value. The energy-selection efficiencies for the trigger and subsequent events were 0.683 and 0.997, respectively. Taking into account the 79.4\% branching intensity of the 6.819-MeV $\alpha$-particle emission from $^{219}$Rn, the $^{215}$Po activity in Crystal~\#4 was determined to be $2.08\pm0.97~\mu$Bq/kg. Owing to the limited statistics, the corresponding energy spectra are not shown in Fig.~\ref{fig:DCSpectrum}. The half-lives and activities obtained by the same analysis for Crystal~\#1 data are summarized in Table~\ref{tab:DCSummary}.

\subsubsection{$^{214}\mathrm{Bi}\rightarrow^{214}\mathrm{Po}\rightarrow^{210}\mathrm{Pb}$ in the $^{238}$U decay chain}
Another decay sequence suitable for DC analysis is
\begin{equation}
\begin{aligned}
&{}^{214}\mathrm{Bi}\left(T_{1/2}=19.71(2)~\mathrm{min}\right)\rightarrow \\
&\qquad {}^{214}\mathrm{Po}\left(T_{1/2}=163.6(3)~\mu\mathrm{s}\right)\rightarrow
{}^{210}\mathrm{Pb},
\end{aligned}
\end{equation}
which occurs in the $^{238}$U decay chain. The $^{214}$Bi undergoes $\beta$ decay with a $Q$ value of 3.269~MeV, followed by the $\alpha$ decay of $^{214}$Po with an energy of 7.687~MeV. The $\beta/\gamma$ events are less suitable as trigger events because of their continuous energy spectrum, which requires a relatively wide energy-selection window and consequently increases the rate of random coincidences. Therefore, the $^{214}$Po $\alpha$ decay was used as the trigger, and preceding $\beta/\gamma$ events were searched for.

\begin{figure*}[htbp]
\centering
\subfloat[$^{216}$Po decay-time distribution.\label{fig:Tdiff232Th}]{%
  \includegraphics[width=0.33\textwidth]{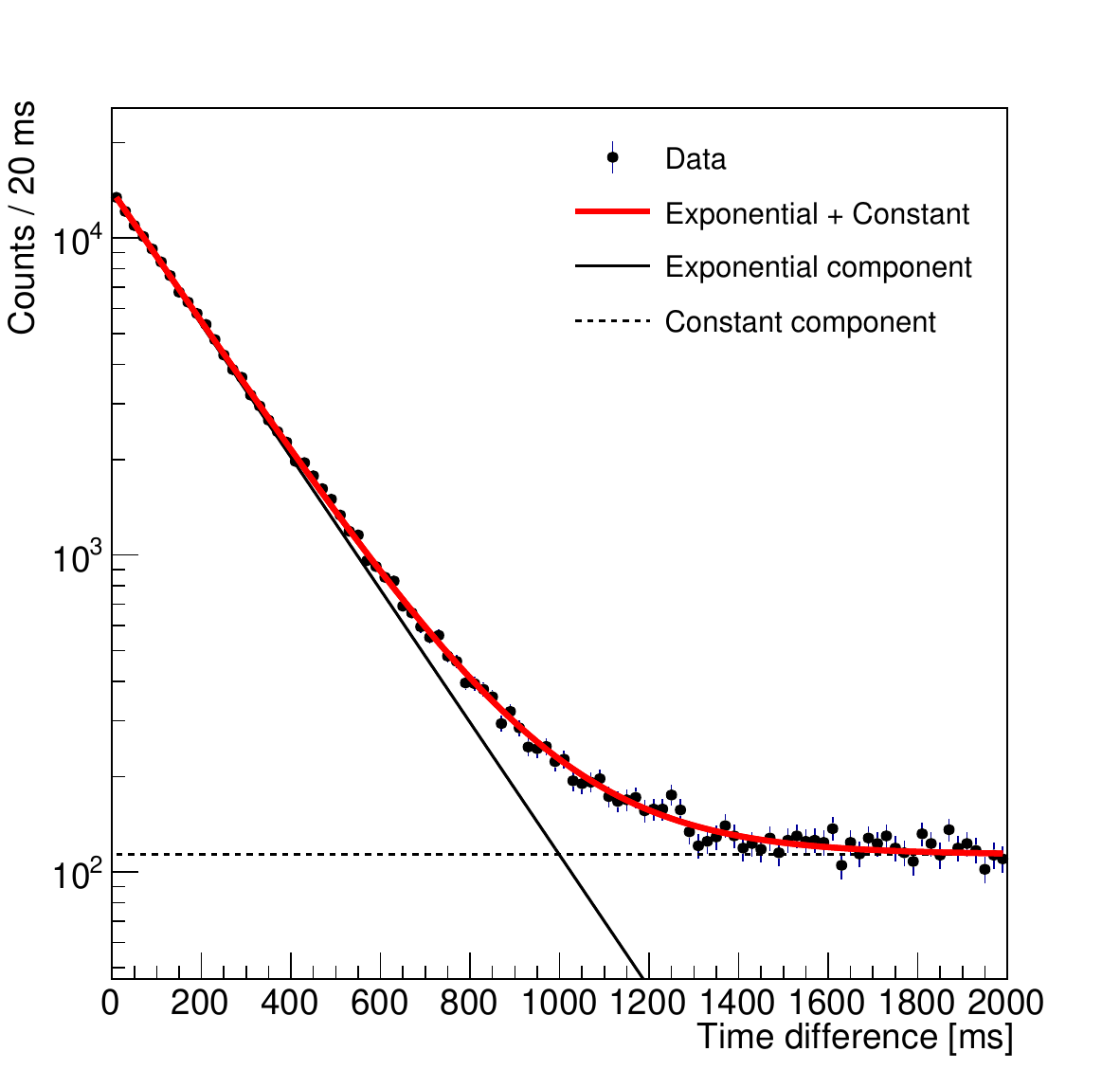}%
}\hspace{1pt}%
\subfloat[$^{215}$Po decay-time distribution.\label{fig:Tdiff235U}]{%
  \includegraphics[width=0.33\textwidth]{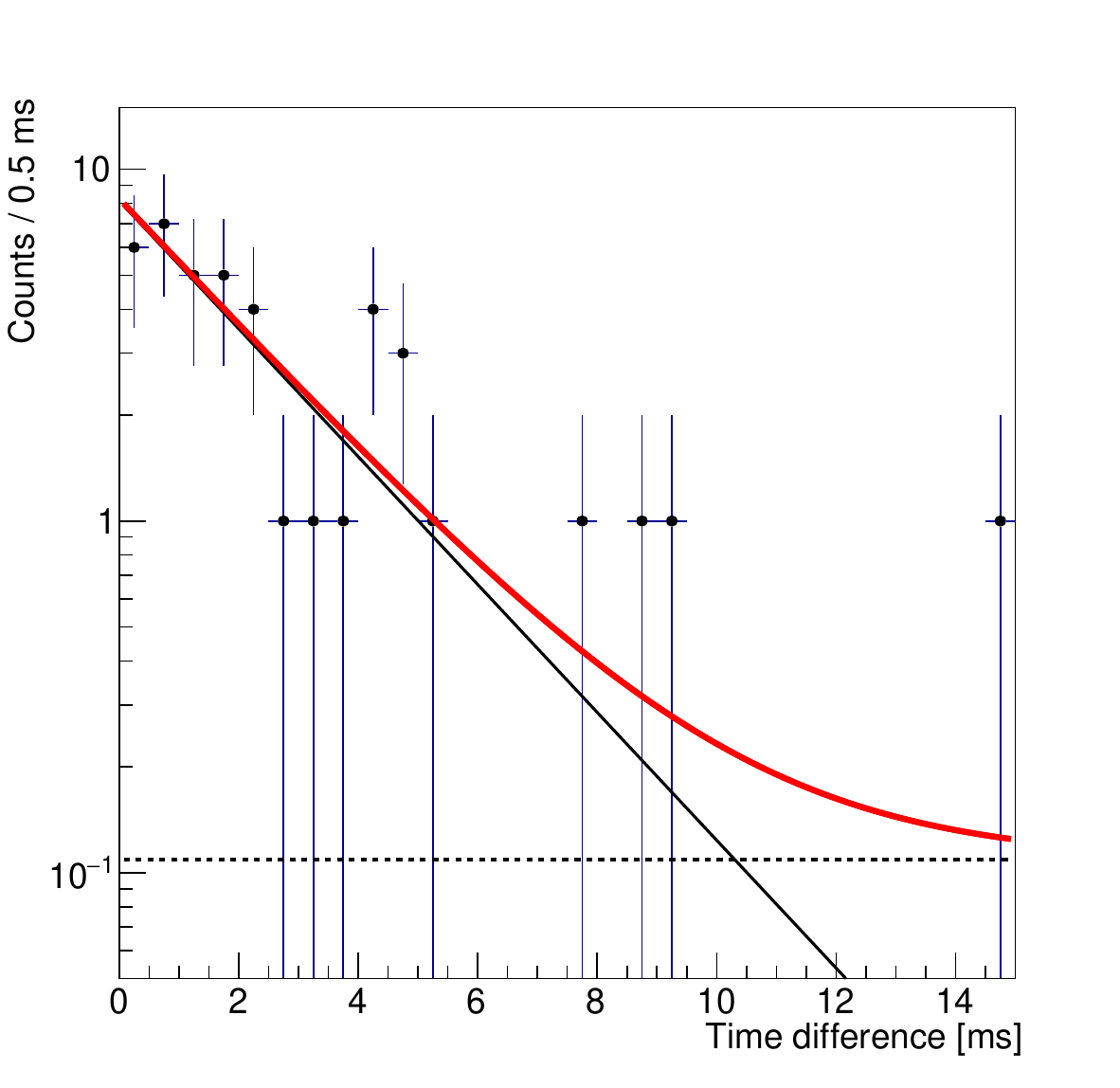}%
}\hspace{1pt}%
\subfloat[$^{214}$Po decay-time distribution.\label{fig:Tdiff238U}]{%
  \includegraphics[width=0.33\textwidth]{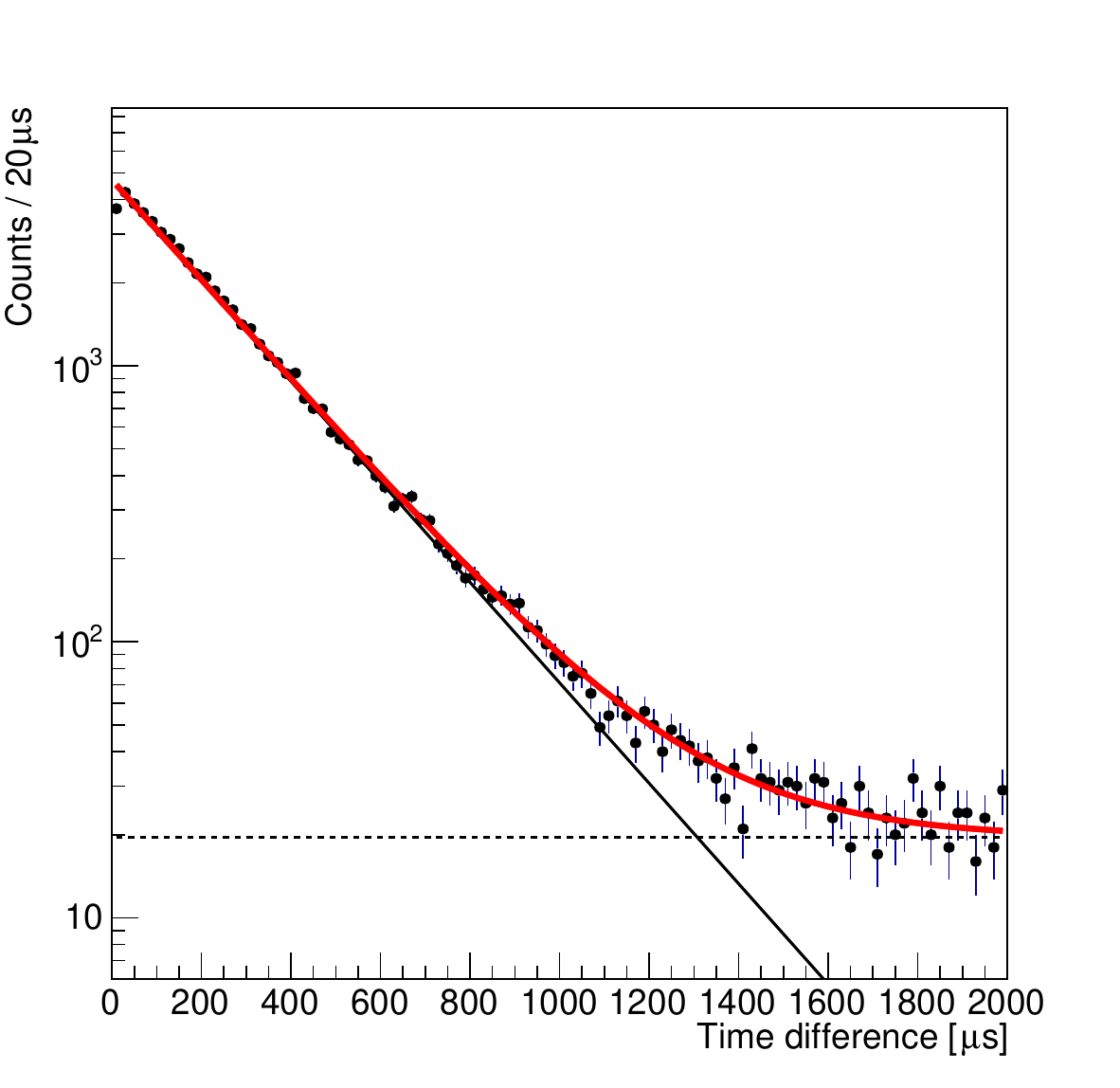}%
}
\caption{Decay-time distributions obtained from the delayed-coincidence analyses of (a) $^{220}$Rn--$^{216}$Po, (b) $^{219}$Rn--$^{215}$Po, and (c) $^{214}$Bi--$^{214}$Po sequential decays in Crystal~\#4. The solid and dashed black lines represent the exponential and constant components, respectively, while the red curves represent their sum.}
\label{fig:DCTimeDistributions}
\end{figure*}

The trigger energy window was defined as $E_{\mathrm{vis}}(7.687)\pm3\sigma_{\alpha}$ for Crystal~\#1 and $E_{\mathrm{vis}}(7.687)\pm15\sigma_{\alpha}$ for Crystal~\#4. The wider window for Crystal~\#4 was adopted because the $^{214}$Po $\alpha$ peak in this crystal deviates significantly from a Gaussian distribution, as discussed below. For each trigger, preceding $\beta/\gamma$ events were searched for within 0--2000~$\mu$s before the trigger. The resulting time-difference distribution for Crystal~\#4 is shown in Fig.~\ref{fig:Tdiff238U}. The half-life obtained from the fit was $165.39\pm0.87~\mu$s, consistent with the reference value. 

\begin{figure}[htbp]
\includegraphics[width=0.45\textwidth]{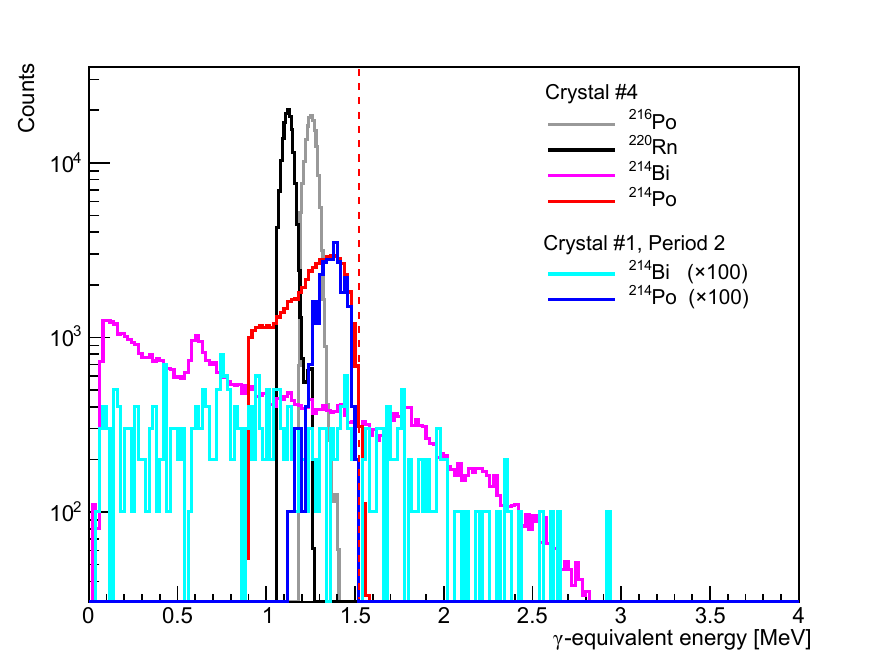}
\caption{\label{fig:DCSpectrum}Energy spectra of the trigger and signal events obtained from the delayed-coincidence (DC) analyses of the $^{232}$Th and $^{238}$U decay chains in Crystal~\#4. The red vertical dashed line indicates the $\gamma$-equivalent energy of the $^{214}$Po $\alpha$-ray, calculated using the quenching function for Crystal~\#4. For comparison, the energy spectra obtained from the DC analysis of the $^{238}$U decay chain in Crystal~\#1 Period~2 are also shown, with their amplitudes multiplied by a factor of 100.}
\end{figure}

The energy spectra of the selected $^{214}$Po and $^{214}$Bi events are shown by the red and magenta lines, respectively, in Fig.~\ref{fig:DCSpectrum}. For comparison, the spectra obtained from the analysis of Crystal~\#1 Period~2 are also shown by the blue and cyan lines, respectively. The $^{214}$Po spectrum for Crystal~\#1 exhibits a single, well-defined peak, whereas that for Crystal~\#4 has a peak position systematically lower than the mean energy of the trigger selection window, indicated by the vertical red dashed line, and a broad tail extending toward lower energies. This anomalous spectral shape suggests that the radioactive contaminants are located near the surface of Crystal~\#4. In contrast, no such anomaly is observed in the $^{214}$Po spectrum of Crystal~\#1, whose surface had been mechanically polished before Phase~1 data taking.

The half-lives and activities obtained from the DC analyses are summarized in Table~\ref{tab:DCSummary}. The half-lives obtained from all datasets are consistent with the corresponding reference values. The activities measured in the two data-taking periods of Crystal~\#1 are also consistent with each other. The evaluated activities are used as constraints and for cross-checks in the spectral analysis described in Sec.~\ref{subsubsec:SpectralFittingProcedure}.

\begin{table*}[htbp]
    \centering
    \begin{threeparttable}
    \caption{
      Summary of half-lives and activities obtained from the delayed-coincidence analysis. The reference half-lives are taken from evaluated nuclear data. The activities are derived from the identified decay pairs assuming secular equilibrium.}
    \label{tab:DCSummary}
    \renewcommand{\arraystretch}{1.15}
    \begin{tabular*}{\textwidth}{@{\extracolsep{\fill}}llccc}
        
        \toprule \toprule
        Parent & Decay sequence & Crystal~\#1 (Period~1) & Crystal~\#1 (Period~2) & Crystal~\#4 \\
        \midrule
        \multirow{2}{*}{$^{228}$Th} & \multirow{2}{*}{$^{220}$Rn\,$\to$\,$^{216}$Po} 
        & $T_{1/2}=144.3\pm2.0$~ms & $T_{1/2}=142.1\pm1.9$~ms & $T_{1/2}=143.8\pm0.5$~ms \\ 
         & & $1.79\pm0.04$~mBq/kg & $1.79\pm0.04$~mBq/kg & $5.37\pm0.03$~mBq/kg \\ 
         \midrule
        \multirow{2}{*}{$^{227}$Ac} & \multirow{2}{*}{$^{219}$Rn\,$\to$\,$^{215}$Po}
        & $T_{1/2}=1.45\pm0.17$~ms & $T_{1/2}=1.71\pm0.21$~ms & $T_{1/2}=1.55\pm0.46$~ms \\
         & & $62.7\pm10.5~\mu$Bq/kg & $69.7\pm11.4~\mu$Bq/kg & $2.08\pm0.97~\mu$Bq/kg \\ \midrule
        \multirow{2}{*}{$^{226}$Ra} & \multirow{2}{*}{$^{214}$Bi\,$\to$\,$^{214}$Po}
        & $T_{1/2}=154.3\pm11.3~\mu$s & $T_{1/2}=171.2\pm20.8~\mu$s & $T_{1/2}=165.4\pm0.9~\mu$s \\
         & & $46.1\pm5.5~\mu$Bq/kg & $49.2\pm7.6~\mu$Bq/kg & $-$\tnote{$\dagger$} \\
        \bottomrule \bottomrule
    \end{tabular*}
    \begin{tablenotes}
        \footnotesize
        \item[$\dagger$] The activity could not be evaluated because of the surface contamination.
    \end{tablenotes}
    \end{threeparttable}
\end{table*}

\subsection{Surface contamination}\label{subsec:SurfaceContamination}
As shown in Fig.~\ref{fig:DCSpectrum}, the low-energy tail of the $^{214}$Po $\alpha$-ray spectrum in Crystal~\#4 provides evidence that the $^{214}$Po contamination is localized near the crystal surface. This feature suggests that the radioactive contaminants are located near the crystal surface, where an insensitive layer is present. In this case, a fraction of the $\alpha$-ray energy is deposited in this layer and does not contribute to the scintillation signal, resulting in a lower observed energy.

A similar interpretation may explain the structures observed around 0.6~MeV and 1.7~MeV in the $^{214}$Bi $\beta/\gamma$ spectrum in Fig.~\ref{fig:DCSpectrum}. If $^{214}$Bi decays occur near the crystal surface, the emitted $\beta$ particle may deposit its energy in the insensitive layer, while an accompanying $\gamma$-ray can penetrate into the scintillating region. Such events can thereby produce structures corresponding to the energies of $\gamma$ rays emitted following $^{214}$Bi decay.

The surface-localization hypothesis is further supported by an independent measurement using a high-purity germanium (HPGe) detector at the Kamioka underground laboratory~\cite{Ichimura2023}. After mechanically polishing the side surface of Crystal~\#4 by approximately 500~$\mu$m, the count rate of the 609-keV $\gamma$-ray from $^{214}$Bi decreased from $8.05\times10^{-3}$~counts/kg/s to $<0.81\times10^{-3}$~counts/kg/s at 90\% C.L. under otherwise identical measurement conditions. This significant reduction by more than an order of magnitude strongly supports the interpretation that the $^{214}$Bi and $^{214}$Po contamination is concentrated near the surface of Crystal~\#4.

To build an accurate background model for Crystal~\#4, the thickness of the insensitive layer and the depth distribution of $^{214}$Bi and $^{214}$Po near the crystal surface were estimated from the shape of the $^{214}$Po $\alpha$ spectrum, as described in Sec.~\ref{subsubsec:SurfaceContaminationModel}.

\subsection{Background modeling}\label{subsec:BackgroundModeling}
The observed energy spectra contain contributions from various radioactive contaminants. To identify and quantify these contributions, background models were constructed based on GEANT4 simulations incorporating the detector responses evaluated in Sec.~\ref{sec:DetectorResponse}. The background sources considered include the internal U/Th decay chains, $^{40}$K, and $^{176}$Lu present in the GAGG scintillator, the external $\gamma$-ray 
backgrounds, and the surface contamination characterized in Sec.~\ref{subsec:SurfaceContamination}. The construction of the background models and the spectral fitting procedure are described in the following subsections.

\subsubsection{Monte Carlo simulation}\label{subsubsec:MonteCarloSimulation}
The background spectra observed by the detector were modeled using a GEANT4-based Monte Carlo simulation. The detailed procedure for constructing the background model, including the simulated detector geometry, radioactive contamination sources, and event generation, is described in Ref.~\cite{Omori2026}. Notably, since the detailed internal structure of the PMT is complex and proprietary, the external backgrounds were modeled under the assumption that the radioactive isotopes were uniformly distributed on the photocathode.

In this work, GEANT4 was used to simulate the energy deposited in the GAGG scintillator by radiations from the background sources. The simulated energy deposits of $\alpha$ rays were converted to the corresponding $\gamma$-equivalent energies using the quenching response shown in Fig.~\ref{fig:Quenching}. Similarly, the energy deposits of $\beta/\gamma$ rays and de-excitation $\gamma$ rays following $\alpha$ decays were corrected using the parameterized nonproportional responses. Finally, the energy resolutions shown in Fig.~\ref{fig:Resolution} were applied to the corrected energies. These procedures yielded the expected energy spectra for the relevant radioactive background components, which were subsequently used as templates in the spectral fitting described in Sec.~\ref{subsubsec:SpectralFittingProcedure}.

\subsubsection{Surface contamination model}\label{subsubsec:SurfaceContaminationModel}
As described in Secs.~\ref{subsec:DelayedCoincidenceAnalysis} and~\ref{subsec:SurfaceContamination}, the DC analysis and measurements with the germanium detector indicated that the $^{214}$Bi and $^{214}$Po contamination is localized near the surface of Crystal~\#4, and that the layer adjacent to the surface exhibits reduced scintillation response. In particular, since this spectral feature persisted throughout the data-taking period, we concluded that it originated from $^{226}$Ra contamination near the crystal surface, rather than transient radon progeny deposition. Therefore, the contribution from the $^{226}$Ra decay chain distributed near the crystal surface needs to be modeled to construct a precise background model. This analysis follows a similar approach to that described in Ref.~\cite{Yu2021}.

We prepared template spectra assuming that $^{214}$Po was located at a depth $d$ from the crystal surface, with a dead layer extending from the surface to a depth $DL$. The dead layer was assumed to be a step-like region with no scintillation response. The continuous variation of the insensitive layer depth is approximated by a superposition of step-function templates with discrete $DL$ values. The template spectra were generated for $d = 0$--$5~\mu$m in steps of $0.5~\mu$m and for $DL = 0$--$8~\mu$m in steps of $1~\mu$m. Here, $DL = 0~\mu$m corresponds to the case without a dead layer, while $DL = 1~\mu$m corresponds to a dead layer extending from the surface to a depth of $1~\mu$m. These templates were generated following the procedure described in Sec.~\ref{subsubsec:MonteCarloSimulation}.

\begin{figure}[htbp]
\includegraphics[width=0.49\textwidth]{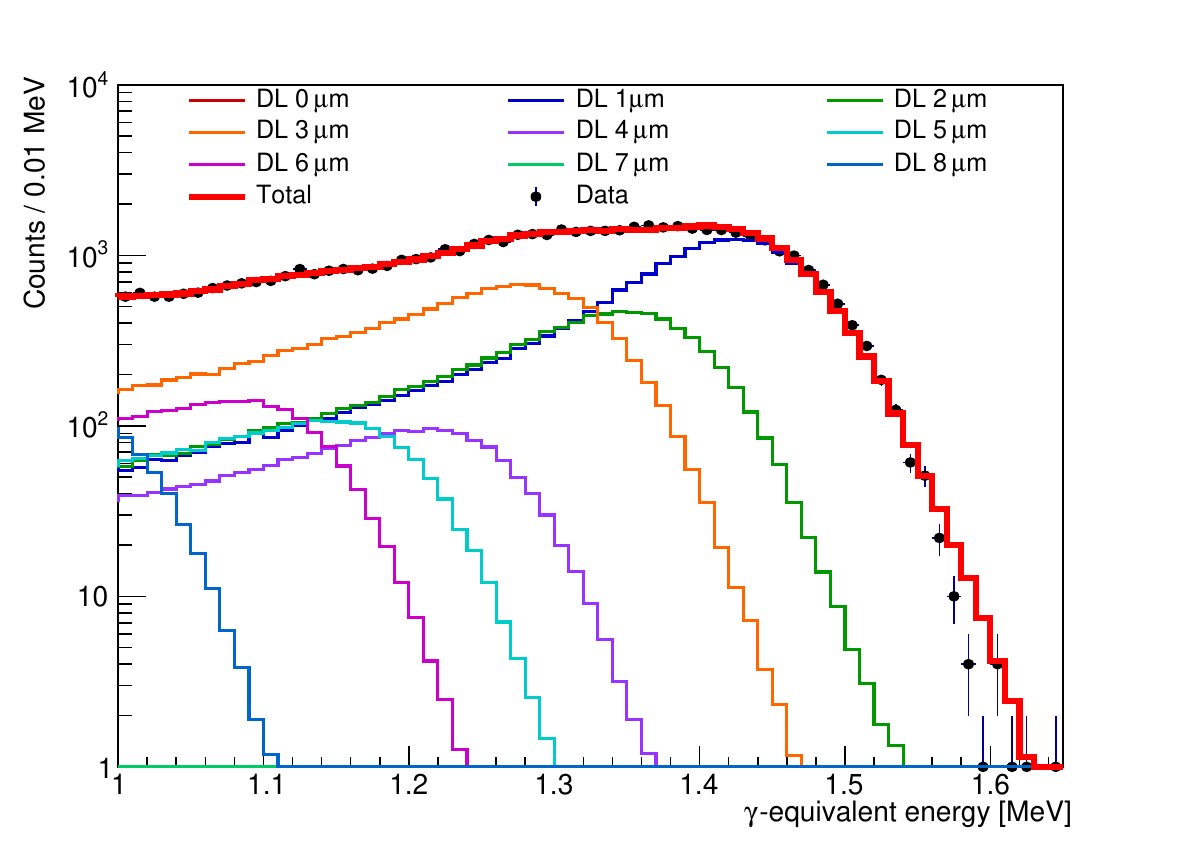}
\caption{
  \label{fig:Po214Surface}
  The $^{214}$Po $\alpha$-ray energy spectrum in Crystal~\#4, selected by the delayed-coincidence analysis, and the combined fit using templates with different dead-layer thicknesses $DL$. The colored lines represent the individual $DL$ components, and the thick red line represents their sum.}
\end{figure}

Assuming that the probability density of $^{214}$Po contamination decreases exponentially with the depth $d$ from the crystal surface with a mean depth $p_1$, the resulting energy distribution $N(E)$ can be expressed as
\begin{equation}
N(E) = p_0 \sum_{DL=0}^{8~\mu \mathrm{m}} F_{DL} \sum_{d=0}^{5~\mu \mathrm{m}} 
N_d^{DL}(E) \exp\!\left(-\frac{d}{p_1}\right),
\label{eq:SurfaceModel}
\end{equation}
where $p_0$ denotes the amplitude, $F_{DL}$ the fraction of events corresponding to each dead-layer thickness $DL$, and $N_d^{DL}(E)$ the energy distribution for a given contamination depth $d$ and dead-layer thickness $DL$. Figure~\ref{fig:Po214Surface} shows the $^{214}$Po $\alpha$ spectrum fit with the combined templates over the energy range of $1.00$--$1.65$~MeV, where the colored curves represent the contributions from the different $DL$ values. The fit yielded a mean depth of $p_1 = 231.9 \pm 5.5$~nm and the following fractions for the dead-layer thicknesses: $F_{0} = 0.0\%$, $F_{1} = 29.4\%$, $F_{2} = 15.1\%$, $F_{3} = 26.6\%$, $F_{4} = 4.6\%$, $F_{5} = 5.9\%$, $F_{6} = 8.8\%$, $F_{7} = 0.0\%$, and $F_{8} = 9.6\%$. Using these parameters, the $^{226}$Ra surface contamination contribution to the $\beta/\gamma$-ray spectrum was modeled and incorporated into the spectral fitting described in Sec.~\ref{subsubsec:SpectralFittingProcedure}.

To obtain the model spectra for the surface $^{226}$Ra contamination, isotopes in the $^{226}$Ra decay sub-chain were generated following the procedure described in Sec.~\ref{subsubsec:MonteCarloSimulation}, assuming an exponential depth distribution with a mean depth $p_1$ for each dead-layer thickness $DL$. The resulting spectra were combined using the fractions $F_{DL}$ determined above. 

In the $\alpha$-ray spectrum of Crystal~\#4, shown in Fig.~\ref{fig:Spectrum}, a non-negligible number of events is observed in the energy region between the $^{152}$Gd/$^{147}$Sm and $^{238}$U peaks. These events can be attributed to incomplete energy deposition of $\alpha$ particles emitted in the decays of $^{238}$U and $^{234}$U. This observation suggests that $^{238}$U and $^{234}$U, which are the dominant internal contaminants, are present not only in the crystal bulk but also near the crystal surface, similarly to $^{226}$Ra. Therefore, spectral templates for the surface $^{238}$U and $^{234}$U contamination were additionally constructed. 

These surface models were incorporated into the fit model for Crystal~\#4 described in Sec.~\ref{subsubsec:SpectralFittingProcedure}.

\subsubsection{Spectral fitting procedure}\label{subsubsec:SpectralFittingProcedure}
The fitting parameters were defined as the activities of internal background sources and the count rates of external backgrounds contributing to the observed energy spectra. For the U/Th decay chains, radioactive equilibrium was assumed based on the half-lives, as listed in Table~\ref{tab:ActivitySummary}, and the activities of the corresponding sub-chains were treated as independent fitting parameters.

The spectral fits were performed by minimizing the Poisson deviance. For each energy bin of the spectrum, the observed number of events was assumed to follow a Poisson distribution, and the corresponding log-likelihood was expressed as the $\chi^2$ statistic~\cite{Baker1984}:
\begin{equation}
\chi^{2}_{\mathrm{energy}}
= 2\sum_i \left[\nu_i - n_i + n_i\ln\!\left(\frac{n_i}{\nu_i}\right)\right],
\label{eq:Chi2Energy}
\end{equation}
where $n_i$ and $\nu_i$ denote the observed and expected numbers of events in the $i$-th energy bin, respectively. For radioactive contaminants with independently measured activities, the measured values and their uncertainties were incorporated into the fit as Gaussian penalty terms:
\begin{equation}
\chi^{2}_{\mathrm{penalty}}
= \sum_{\mu} \left(\frac{A_{\mu} - A_{\mu}^{\mathrm{ref}}}
{\sigma_{\mu}^{\mathrm{ref}}}\right)^2,
\label{eq:Chi2Penalty}
\end{equation}
where $A_{\mu}$ denotes the fitted activity of the $\mu$-th radioactive contaminant, and $A_{\mu}^{\mathrm{ref}}$ and $\sigma_{\mu}^{\mathrm{ref}}$ denote the reference activity and its uncertainty from an independent measurement, respectively. The total fitting statistic was then defined as

\begin{equation}
\chi^{2}_{\mathrm{total}}
= \chi^{2}_{\mathrm{energy}} + \chi^{2}_{\mathrm{penalty}}.
\label{eq:Chi2Total}
\end{equation}

In this study, physical non-negativity constraints ($A_i \geq 0$) were imposed on all background parameters throughout the fitting procedure. The spectral fits were performed sequentially for each dataset: the $\alpha$-ray spectrum was first fit to the internal U/Th background models, and the resulting activities were subsequently used as constraints in the $\beta/\gamma$-ray spectral fit.

For Crystal~\#1, the data were divided into Period~1 and Period~2 as described in Sec.~\ref{subsec:DataAcquisition}, and the spectral fits were performed independently for each period. Since no significant change in bulk radioactive contamination is expected between the two periods, the activities of the internal background sources were treated as common parameters shared between Period~1 and Period~2. The $\alpha$-ray spectra were fit over the energy range of $0.50$--$1.20$~MeV. The $^{215}$Po and $^{214}$Po activities of $65.9 \pm 7.7~\mu$Bq/kg and $47.2 \pm 4.5~\mu$Bq/kg, respectively, obtained as inverse-variance-weighted mean values from the DC analysis of the two datasets (Sec.~\ref{subsec:DelayedCoincidenceAnalysis}), were applied as penalty constraints on the $^{227}$Ac and $^{226}$Ra contributions, with the activities expressed in radioactive-equivalent units of the respective decay sub-chains. 

Subsequently, the $\beta/\gamma$-ray spectra were fit over the energy range of $0.86$--$2.70$~MeV. The lower bound was chosen as the energy below which the background model no longer adequately reproduces the observed spectrum; the discrepancy at lower energies is likely attributable to low-energy Compton scattering contributions from external $\gamma$-ray sources not fully characterized in the present simulation. A detailed investigation of this region is left for future work. The fitting parameters are the activities of internal $^{40}$K, $^{176}$Lu, internal U/Th decay chains, and external $^{40}$K and U/Th sources. The U/Th chain activities obtained from the $\alpha$-ray fit were applied as penalty constraints. In addition, the $^{40}$K and $^{176}$Lu activities measured with the HPGe detector, $<3.69$~mBq/kg (68\% C.L.) and $5.69 \pm 0.79$ mBq/kg, respectively, were incorporated as penalty constraints. For $^{40}$K, the upper limit was implemented by setting the reference activity $A^{\mathrm{ref}}$ to zero and its uncertainty $\sigma^{\mathrm{ref}}$ to 3.69 mBq/kg.

For Crystal~\#4, the $\alpha$-ray spectra were fit over the energy range of $0.50$--$1.50$~MeV. The $^{227}$Ac activity was constrained to 
$2.1 \pm 1.0~\mu$Bq/kg from the DC analysis of the $^{219}$Rn--$^{215}$Po sequence. The $^{214}$Bi--$^{214}$Po DC analysis was not used to constrain the $^{226}$Ra activity, because the $^{214}$Po $\alpha$-ray spectrum was distorted by the surface contamination as described in Sec.~\ref{subsec:SurfaceContamination}. Instead, additional fitting parameters were introduced to account for the surface contamination characterized in Sec.~\ref{subsubsec:SurfaceContaminationModel}: the activities of the surface $^{226}$Ra, $^{238}$U, and $^{234}$U components. To reduce the number of free parameters, the ratio of the surface $^{238}$U to $^{234}$U activity was constrained to equal that of the corresponding bulk activities determined from the same fit. 

The $\beta/\gamma$-ray spectra were then fit over the same energy range as Crystal~\#1. The activities of the surface $^{226}$Ra and $^{238}$U obtained from the $\alpha$-ray fit were applied as penalty constraints, in addition to the U/Th bulk activities. The activities of $^{40}$K and $^{176}$Lu were constrained to $<1.59$~mBq/kg (68\% C.L.) and $1.38 \pm 0.31$~mBq/kg, respectively, as determined by HPGe measurement.

\section{Results}\label{sec:Results}
\begin{figure*}[t]
    \centering
    \setlength{\abovecaptionskip}{2pt}
    \setlength{\belowcaptionskip}{0pt}
    \captionsetup[subfloat]{labelformat=empty}

    \subfloat[\parbox{0.44\textwidth}{\centering\footnotesize (a) $\alpha$-ray spectrum, Crystal~\#1, Period~1\\$\chi^2/\mathrm{ndf}=945.4/340$}]{%
        \includegraphics[width=0.45\textwidth, trim=0mm 5mm 10mm 10mm, clip]{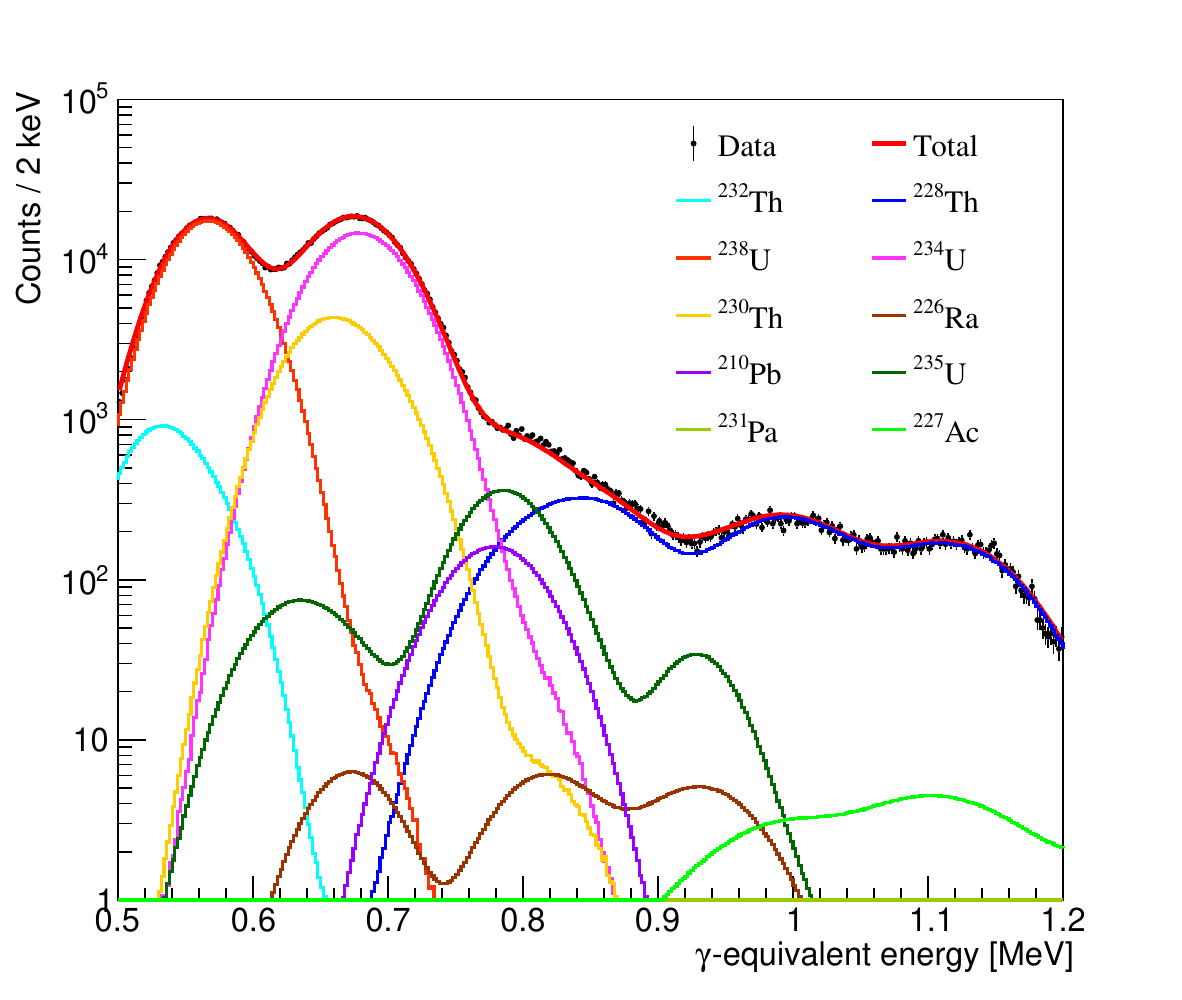}
        \label{fig:AlphaFitCrystal1Period1}
    }
    \hfill
    \subfloat[\parbox{0.44\textwidth}{\centering\footnotesize (b) $\beta/\gamma$-ray spectrum, Crystal~\#1, Period~1\\$\chi^2/\mathrm{ndf}=256.6/170$}]{%
        \includegraphics[width=0.45\textwidth, trim=0mm 5mm 10mm 10mm, clip]{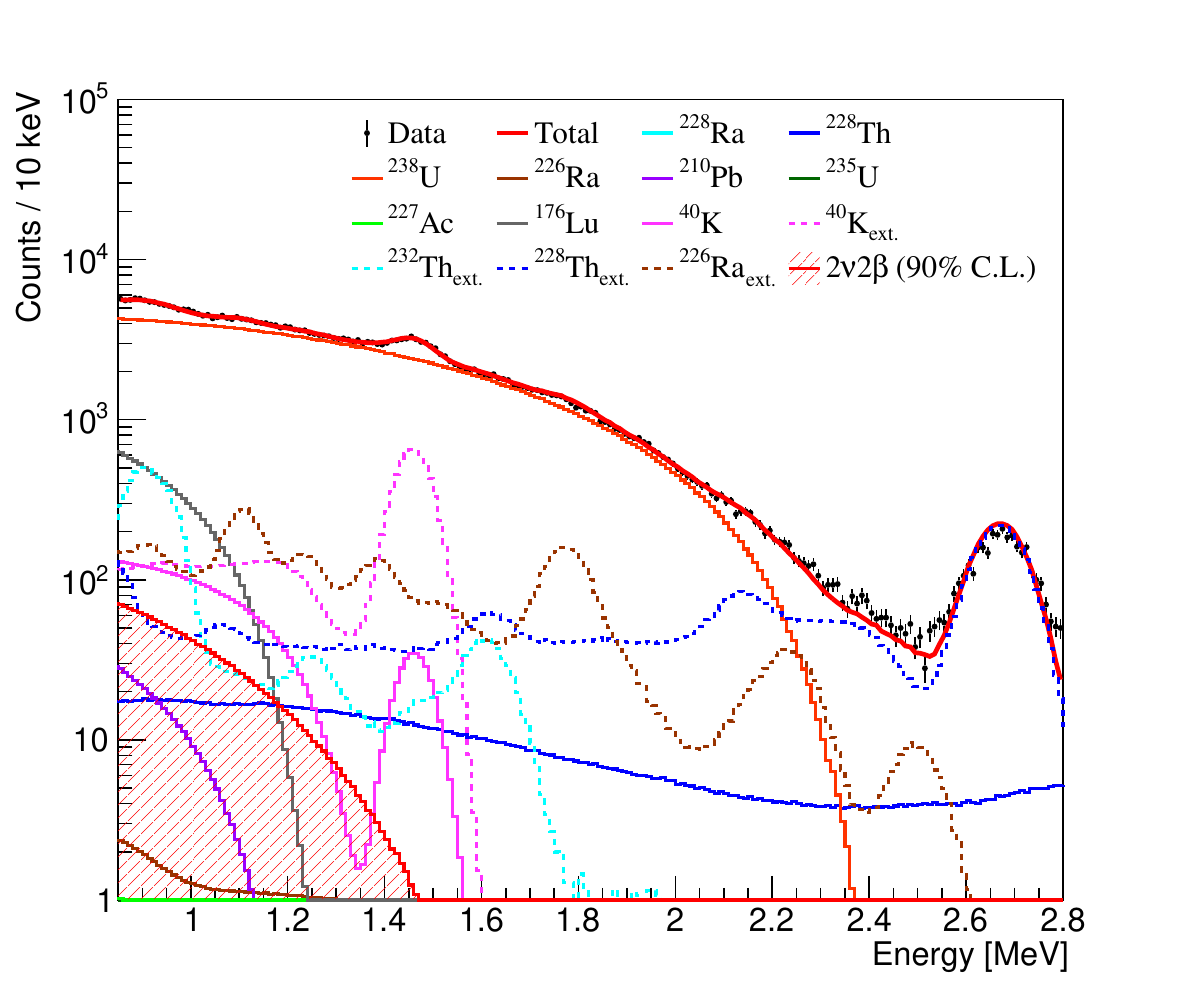}
        \label{fig:BetaFitCrystal1Period1}
    }

    \vspace{-3mm}

    \subfloat[\parbox{0.44\textwidth}{\centering\footnotesize (c) $\alpha$-ray spectrum, Crystal~\#1, Period~2\\$\chi^2/\mathrm{ndf}=829.6/340$}]{%
        \includegraphics[width=0.45\textwidth, trim=0mm 5mm 10mm 10mm, clip]{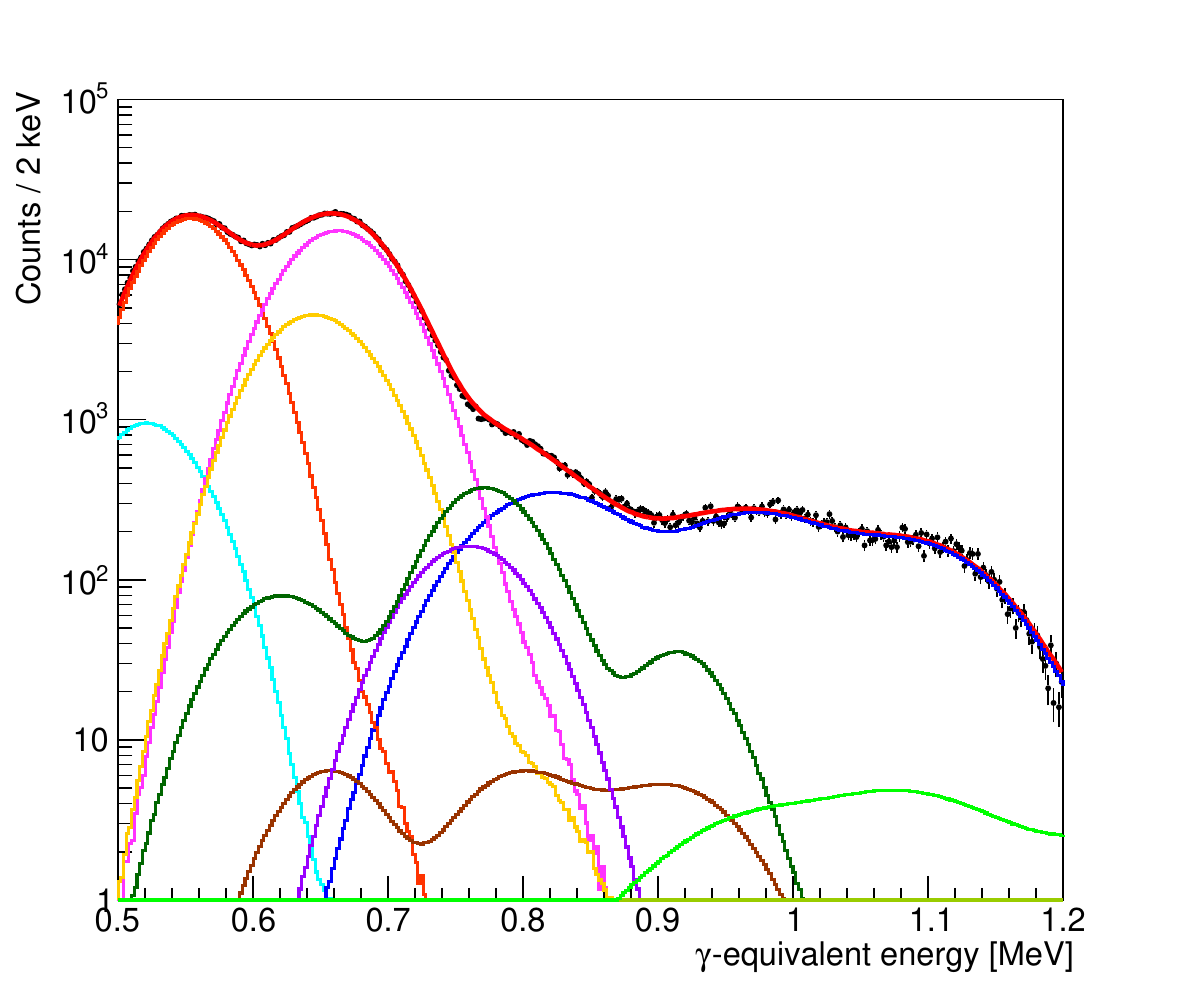}
        \label{fig:AlphaFitCrystal1Period2}
    }
    \hfill
    \subfloat[\parbox{0.44\textwidth}{\centering\footnotesize (d) $\beta/\gamma$-ray spectrum, Crystal~\#1, Period~2\\$\chi^2/\mathrm{ndf}=257.9/170$}]{%
        \includegraphics[width=0.45\textwidth, trim=0mm 5mm 10mm 10mm, clip]{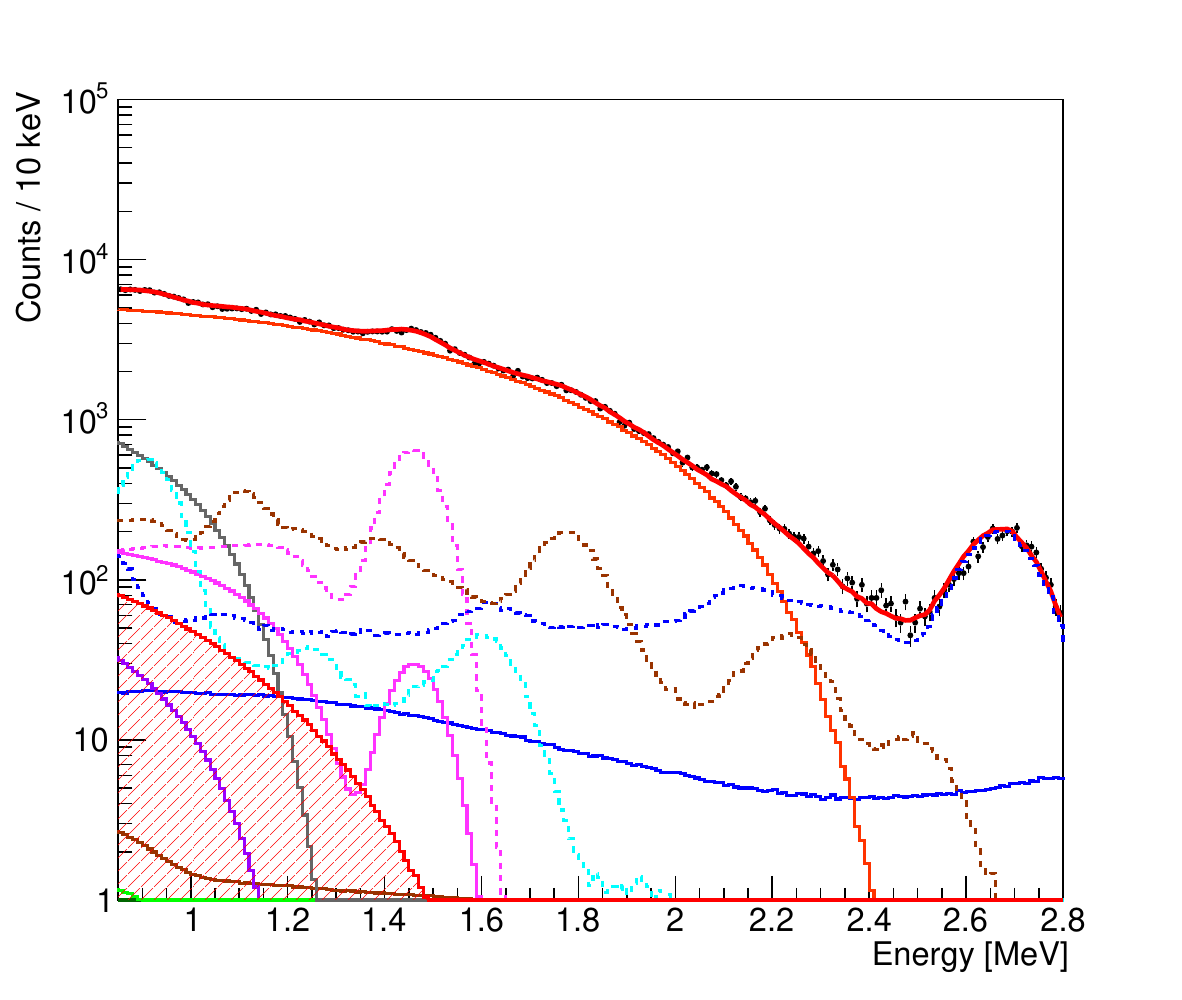}
        \label{fig:BetaFitCrystal1Period2}
    }

    \vspace{-3mm}

    \subfloat[\parbox{0.44\textwidth}{\centering\footnotesize (e) $\alpha$-ray spectrum, Crystal~\#4\\$\chi^2/\mathrm{ndf}=11582.7/488$}]{%
        \includegraphics[width=0.45\textwidth, trim=0mm 5mm 10mm 10mm, clip]{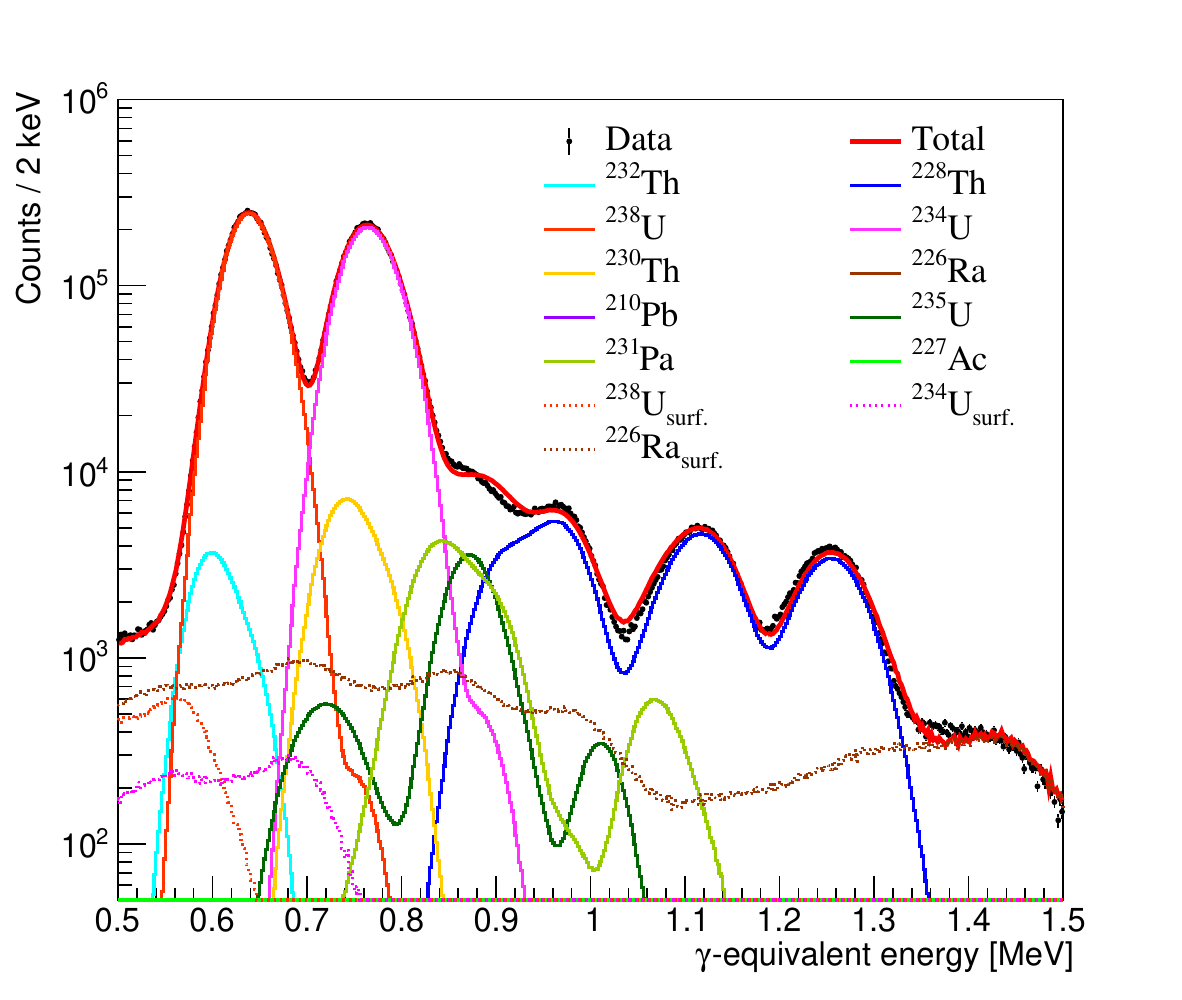}
        \label{fig:AlphaFitCrystal4}
    }
    \hfill
    \subfloat[\parbox{0.44\textwidth}{\centering\footnotesize (f) $\beta/\gamma$-ray spectrum, Crystal~\#4\\$\chi^2/\mathrm{ndf}=554.0/168$}]{%
        \includegraphics[width=0.45\textwidth, trim=0mm 5mm 10mm 10mm, clip]{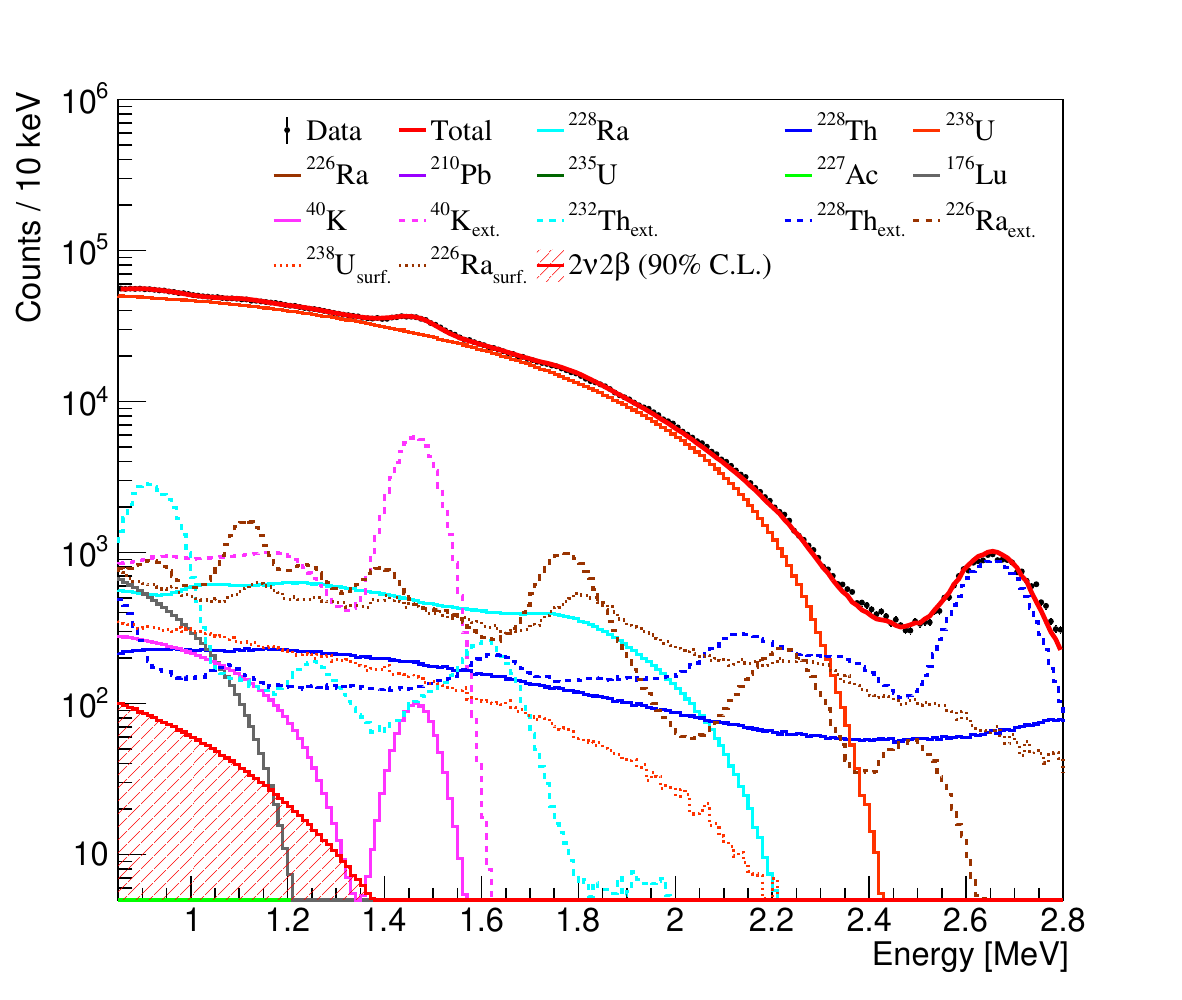}
        \label{fig:BetaFitCrystal4}
    }

    \caption{
      Best-fit results for all datasets. The thick red lines represent the total background model, while the colored solid, dotted, and dashed lines represent the internal, surface, and external background contributions, respectively. The $^{160}$Gd $2\nu2\beta$ signal spectrum at the 90\% C.L. lower limit derived in Sec.~\ref{sec:Results} is overlaid on the $\beta/\gamma$-ray spectra for illustration only.}
    \label{fig:FittingResult}
\end{figure*}

Figure~\ref{fig:FittingResult} shows the reproduced $\alpha$- and $\beta/\gamma$-ray spectra for all datasets using the modeled background components. The thick red lines represent the total background model, while the colored solid, dotted, and dashed lines represent the contributions from internal, surface, and external backgrounds, respectively. The external and surface background components are denoted with the subscripts ``ext'' and ``surf,'' respectively. For illustration, the $^{160}$Gd $2\nu2\beta$ signal spectrum corresponding to the 90\% C.L. lower limit derived in Sec.~\ref{sec:Results} is overlaid on the $\beta/\gamma$-ray spectra. The fitted activities are summarized in Table~\ref{tab:ActivitySummary}. Most of the fitted activities are consistent with the results of the DC analysis and HPGe measurement. Although the fitted $^{227}$Ac activity in Crystal~\#1 differs from the result of the DC analysis, this isotope constitutes a minor background component, with an activity on the order of $\mu$Bq/kg, and has little impact on the reproduction of the observed energy spectra.
 
For Crystal~\#4, the surface components play a critical role in reproducing both the shoulder structure around 1.4~MeV and low-energy tail below 0.55~MeV (Fig.~\ref{fig:AlphaFitCrystal4}). This demonstrates that the surface contamination model developed in Sec.~\ref{subsubsec:SurfaceContaminationModel} is essential for an adequate description of the Crystal~\#4 data.

\begin{table*}[htbp]
\centering
\caption{
  Summary of the activities or detection rates obtained from the $\alpha$- and $\beta/\gamma$-ray spectral fits. The activities of the internal backgrounds are given in units of mBq/kg, while those of the external backgrounds are given in units of $\times10^{-3}$~counts/kg/s. The quoted uncertainties are statistical only. Upper limits are at 90\% C.L.
}
\label{tab:ActivitySummary}

\begin{tabular*}{0.70\textwidth}{@{\extracolsep{\fill}}llccc}
\toprule
\toprule
\multicolumn{5}{c}{\textbf{Internal backgrounds [mBq/kg]}} \\
\midrule
Chain & Parent & \multicolumn{2}{c}{Crystal~\#1} & Crystal~\#4 \\
\midrule

\multirow{3}{*}{$^{232}$Th}
 & $^{232}$Th  & \multicolumn{2}{c}{$6.4\pm0.1$}   & $4.3\pm0.1$ \\
 & $^{228}$Ra  & \multicolumn{2}{c}{$<9.2\times10^{-5}$} & $3.8\pm0.3$ \\
 & $^{228}$Th  & \multicolumn{2}{c}{$1.88\pm0.01$}  & $5.55\pm0.01$ \\
\midrule

\multirow{8}{*}{$^{238}$U}
 & $^{238}$U                    & \multicolumn{2}{c}{$119.7\pm0.1$}  & $274.1\pm0.1$ \\
 & $^{238}$U$_{\rm surf}$       & \multicolumn{2}{c}{$-$}            & $5.1\pm0.1$ \\
 & $^{234}$U                    & \multicolumn{2}{c}{$117.6\pm0.4$}  & $265.4\pm0.2$ \\
 & $^{234}$U$_{\rm surf}$       & \multicolumn{2}{c}{$-$}            & $4.9\pm 0.1$ \\
 & $^{230}$Th                   & \multicolumn{2}{c}{$35.6\pm0.4$}   & $9.5\pm0.2$ \\
 & $^{226}$Ra                   & \multicolumn{2}{c}{$(49.4\pm4.5)\times10^{-3}$ } & $<1.9\times10^{-4}$ \\
 & $^{226}$Ra$_{\rm surf}$      & \multicolumn{2}{c}{$-$}            & $7.41\pm0.05$ \\
 & $^{210}$Pb                   & \multicolumn{2}{c}{$1.3\pm0.2$}    & $<2.8\times10^{-3}$ \\
\midrule

\multirow{3}{*}{$^{235}$U}
 & $^{235}$U  & \multicolumn{2}{c}{$3.7\pm0.2$}             & $5.7\pm0.1$ \\
 & $^{231}$Pa & \multicolumn{2}{c}{$<28.5\times10^{-3}$}    & $8.3\pm0.1$ \\
 & $^{227}$Ac & \multicolumn{2}{c}{$(25.2\pm6.2)\times10^{-3}$} & $(3.5\pm1.0)\times10^{-3}$ \\
\midrule

 & $^{176}$Lu & \multicolumn{2}{c}{$6.2\pm0.6$}  & $1.6\pm0.3$ \\
 & $^{40}$K   & \multicolumn{2}{c}{$3.0\pm2.2$}  & $<2.11$ \\

\midrule
\midrule
\multicolumn{5}{c}{\textbf{External backgrounds [$\times10^{-3}$ counts/kg/s]}} \\
 & Parent & \multicolumn{2}{c}{Crystal~\#1} & Crystal~\#4 \\
\midrule

 &  & Period~1 & Period~2 & \\
\cmidrule{3-4}
 & $^{40}$K$_{\rm ext}$    & $6.4\pm0.3$ & $7.3\pm 0.3$ & $10.2\pm0.2$ \\
 & $^{232}$Th$_{\rm ext}$  & $11.4\pm1.2$ & $12.5\pm 1.2$ & $11.4\pm0.7$ \\
 & $^{228}$Th$_{\rm ext}$  & $18.2\pm0.3$ & $19.8\pm 0.4$ & $12.6\pm0.1$ \\
 & $^{226}$Ra$_{\rm ext}$  & $20.1\pm1.1$ & $26.9\pm 1.2$ & $21.2\pm0.8$ \\

\bottomrule
\bottomrule
\end{tabular*}
\end{table*}

The $^{160}$Gd $2\nu2\beta$ signal spectrum, modeled in Ref.~\cite{Omori2026}, was also included as a free parameter in the $\beta/\gamma$-ray spectral fits. For Crystal~\#1, the signal rate was treated as a common parameter across Periods~1 and~2. The signal rates were found to be $(-3.79\pm2.91)\times10^{-3}$~counts/kg/s and $(-2.92\pm1.79)\times10^{-3}$~counts/kg/s for Crystal~\#1 and Crystal~\#4, respectively, consistent with zero within the statistical uncertainty. No statistically significant excess attributable to $^{160}$Gd $2\nu2\beta$ decay was observed.

An upper limit on the signal count $S_{\rm lim}$ can be used to derive a lower limit on the half-life through
\begin{equation}
T_{1/2} \geq \frac{\ln 2 \, N \varepsilon t}{S_{\rm lim}},
\label{eq:Limit}
\end{equation}
where $N$ is the number of $^{160}$Gd nuclei in the GAGG scintillator, $\varepsilon$ is the detection efficiency for the $2\nu2\beta$ decay, $t$ is the live time of the measurement, and $S_{\rm lim}$ is the upper limit on the number of signal events at a given confidence level. 

In this analysis, the effects of electron escape and bremsstrahlung-photon emission are included in the GEANT4 simulation and are therefore already reflected in the signal template. Consequently, $\varepsilon = 1$ effectively, and no additional efficiency correction is required. The number of $^{160}$Gd nuclei is given by
\begin{equation}
N = \frac{M a N_A}{W},
\end{equation}
where $M = 0.383$~kg and $1.24$~kg are the masses of Gd contained in Crystal~\#1 and Crystal~\#4, respectively, $a$ is the natural isotopic abundance of $^{160}$Gd, $W$ is the molar mass of Gd, and $N_A$ is Avogadro's constant, yielding $N = 3.22\times10^{23}$ and $1.04\times10^{24}$ for Crystal~\#1 and Crystal~\#4, respectively.

The upper limit on the signal count $S_{\rm lim}$ was estimated using a profile likelihood method applied to the $2\nu2\beta$ signal parameter. The resulting $S_{\rm lim}$ at 90\% (68\%) C.L. are 19015 (10220) events and 12582 (5094) events for Crystal~\#1 and Crystal~\#4, respectively. These correspond to lower limits on the half-life of $T_{1/2} \geq 5.58\,(10.4)\times10^{18}$~yr and $T_{1/2} \geq 2.03\,(5.00)\times10^{19}$~yr at 90\% (68\%) C.L. for Crystal~\#1 and Crystal~\#4, respectively. The likelihood functions for Crystal~\#1 and Crystal~\#4 were combined under the assumption of a common signal rate, yielding the combined lower limit:
\begin{equation}
T_{1/2} \geq 2.32\,(5.78)\times10^{19}~\mathrm{yr}
\quad \text{at 90\%\,(68\%) C.L.,}
\label{eq:HalfLifeLimit}
\end{equation}
representing the first result of the PIKACHU Phase~1 search for $2\nu2\beta$ decay of $^{160}$Gd to the ground state of $^{160}$Dy.

\section{Discussion}\label{sec:Discussion}
Through the background characterization presented in this paper, the Phase~1 experiment has set a lower limit on the $^{160}$Gd $2\nu2\beta$ half-life comparable to that obtained with a GSO scintillator~\cite{Danevich2001}. In addition, this study has provided new insights into the signal and background correlations that are key to setting more stringent constraints on the $^{160}$Gd $2\nu2\beta$ decay rate in Phase~1. This section discusses the issue, as well as the requirements for proceeding to Phase~2 experiment, which aims to achieve the discovery of $2\nu2\beta$ decay.

\subsection{Correlations between signal and background parameters}\label{subsec:CorrelationsBetweenSignalAndBackgroundParameters}
Assuming that the observed data consist only of background events, $S_{\rm lim}$ in Eq.~\eqref{eq:Limit} can be estimated as $\sqrt{B}$, where $B$ is the number of background events in the analysis energy window of $0.86$--$2.70$~MeV, under Poisson statistics. Using $S_{\rm lim} = 1369$ and $3221$ for Crystals~\#1 and \#4, respectively, together with a signal efficiency of 0.17 within the $\sqrt{B}$ counting energy window, as estimated by a GEANT4 simulation, the expected sensitivities at 90\% C.L. are $T_{1/2} = 1.30\times10^{19}$~yr and $T_{1/2} = 1.35\times10^{19}$~yr for Crystal~\#1 and Crystal~\#4, respectively. While the lower limit obtained for Crystal~\#4 slightly exceeds the expected sensitivity, that for Crystal~\#1 falls short of it. 

\begin{figure}[htbp]
\includegraphics[width=0.45\textwidth]{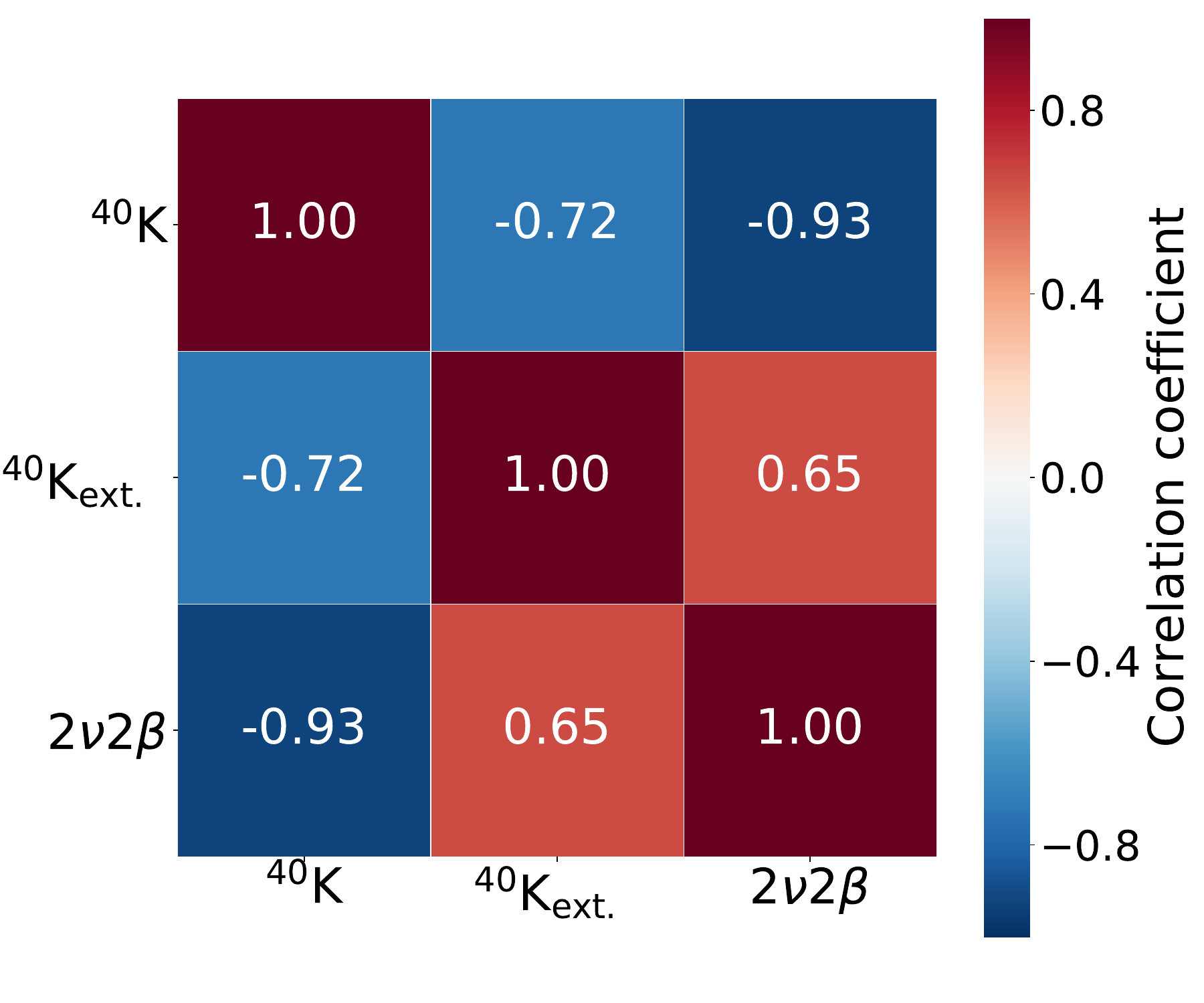}
\caption{\label{fig:HeatMap} Heat map of the correlation coefficients among the internal $^{40}\mathrm{K}$ activity, external $^{40}\mathrm{K}$ rate, and $2\nu2\beta$ signal, obtained from the covariance matrix of the spectral fit for Crystal~\#1 in Period~1.}
\end{figure}

The primary cause of the degradation in Crystal~\#1 is the correlation between the $2\nu2\beta$ signal and internal $^{176}$Lu and $^{40}$K. While $^{176}$Lu negatively correlates with the signal due to its similar spectral shape, the anti-correlation with internal $^{40}$K arises from the following mechanism. The 1.46~MeV $\gamma$-ray peak of internal $^{40}$K is largely absorbed by the external $^{40}$K component in the fit, since both contribute to the same spectral feature. Consequently, the internal $^{40}$K activity is constrained primarily through its $\beta$-decay continuum, which significantly overlaps with the $2\nu2\beta$ signal spectrum. This leads to a three-way degeneracy among the signal rate, the internal $^{40}$K activity, and the external $^{40}$K activity, thereby limiting the sensitivity of the half-life measurement. This interpretation is supported by the strong correlations shown in Fig.~\ref{fig:HeatMap}: the internal $^{40}\mathrm{K}$ activity is negatively correlated with the $2\nu2\beta$ signal, whereas the external $^{40}\mathrm{K}$ rate exhibits a positive correlation with it.

To eliminate $^{176}$Lu contamination and thereby resolve its correlation with the signal, future crystals will be grown using a dedicated new crucible. The crystals used in this work were grown in a crucible previously used for Lu$_3$Al$_5$O$_{12}$ crystal growth, and lutetium diffused into the crucible material during prior use was subsequently introduced into the GAGG crystals during the growth process.

\begin{figure}[htbp]
\includegraphics[width=0.45\textwidth]{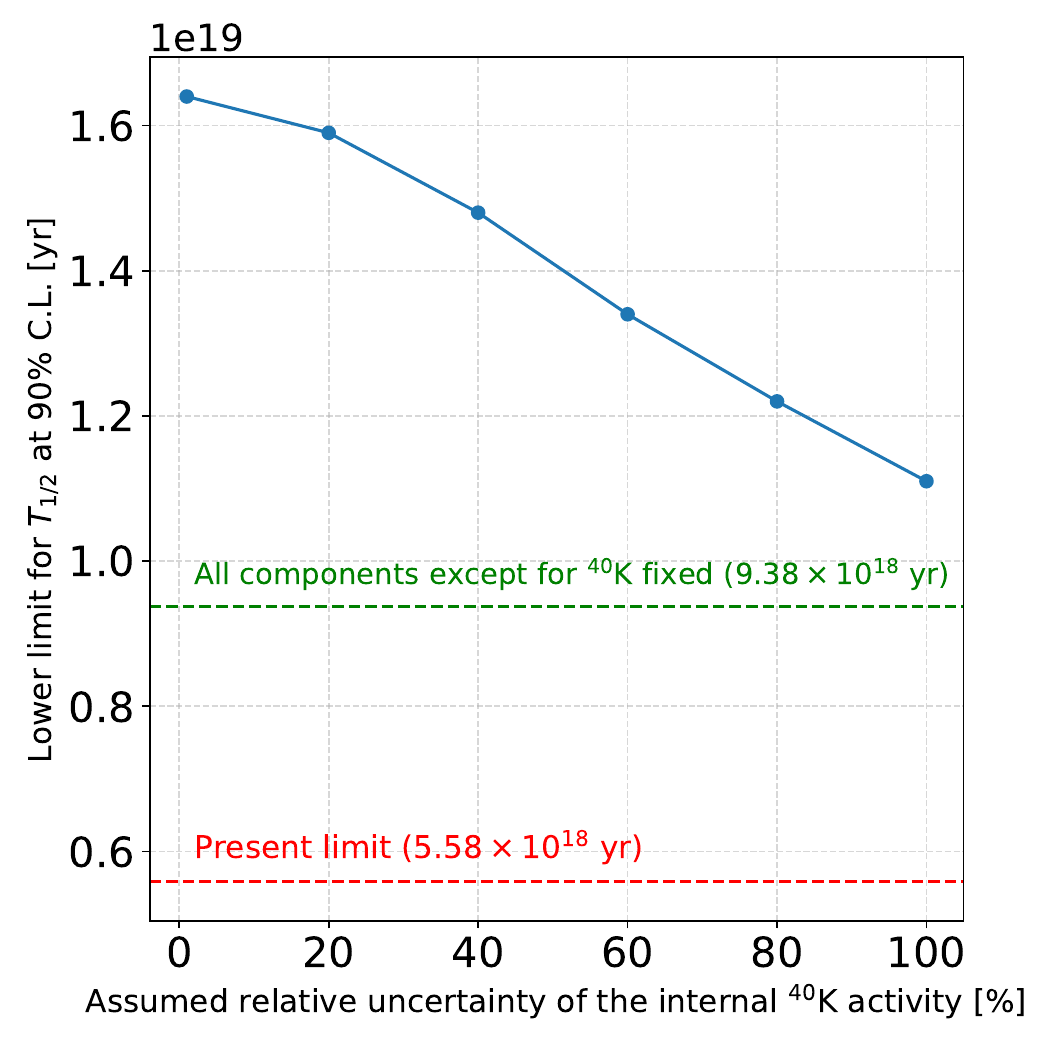}
\caption{\label{fig:Correlation}
    The 90\% C.L. lower limit on the $^{160}$Gd $2\nu2\beta$ half-life for Crystal~\#1 as a function of the assumed precision of an independent measurement to constrain the $^{40}$K activity. The red dashed line indicates the present limit shown in Sec.~\ref{sec:Results}. The green dashed line represents the lower limit calculated with all background components except $^{40}$K fixed at their best-fit values. The blue markers show the lower limits calculated under the same assumption, while constraining the internal $^{40}$K activity to its best-fit value using a Gaussian penalty term with a width corresponding to the precision indicated on the $x$-axis.}
\end{figure}

In contrast, $^{40}$K is ubiquitous in the environment and cannot be easily eliminated from the crystal; its activity must therefore be constrained by an independent measurement. Figure~\ref{fig:Correlation} shows the 90\% C.L. lower limit on the $^{160}$Gd $2\nu2\beta$ half-life for Crystal~\#1 as a function of the assumed precision of an independent measurement of the internal $^{40}$K activity. The red dashed line indicates the present limit, while the green dashed line represents the limit obtained when all background components except $^{40}$K are fixed at their best-fit values, reflecting the contribution of improved external background characterization to the limit for half-life. The blue markers in Fig.~\ref{fig:Correlation} show the limits obtained by repeating the profile-likelihood analysis on the Crystal~\#1 data while varying the uncertainty assigned to the Gaussian penalty term for the internal $^{40}$K activity in Eq.~\eqref{eq:Chi2Penalty}. Specifically, $\sigma^{\mathrm{ref}}$ was set to the precision indicated on the $x$-axis, while $A^{\mathrm{ref}}$ was fixed at its best-fit value. The figure demonstrates that both a precise determination of the internal $^{40}$K activity and an accurate characterization of the remaining background components are effective strategies for enhancing the half-life sensitivity in Phase~1.

In this study, an HPGe measurement conducted over 17~days was insufficient to determine a finite internal $^{40}$K activity in Crystal~\#1. We are therefore currently developing an Inductively Coupled Plasma Mass Spectrometry (ICP-MS) technique to measure the internal $^{40}$K concentration directly. Furthermore, the correlation mechanism described above implies that reducing the external $^{40}$K contribution is also effective in mitigating the correlation between internal $^{40}$K and the signal. This can be achieved through the use of low-background photodetectors, as introduced in Sec.~\ref{subsec:ProspectsForFuturePIKACHUPhase1}.

Although these correlations are also present in Crystal~\#4, the anti-correlated internal $^{40}$K activity reaches its physical boundary at zero before the signal best-fit value shifts into the positive region. Once $^{40}$K is pinned at this boundary, the anti-correlation is relieved and the likelihood rises more steeply with increasing signal, resulting in a tighter upper limit on the signal count. This boundary effect explains why the Crystal~\#4 data achieves a lower limit slightly better than the naive sensitivity estimate.

\subsection{Prospects for future PIKACHU Phase~1}\label{subsec:ProspectsForFuturePIKACHUPhase1}
Besides the approaches to mitigate the parameter correlations described above, hardware upgrades are planned to further improve the experimental sensitivity.

As shown in Table~\ref{tab:ActivitySummary}, the dominant background is the internal $^{238}$U decay chain, in which $^{234\mathrm{m}}$Pa undergoes $\beta$-decay with $Q_{\beta} = 2.27$~MeV. As discussed in Ref.~\cite{Omori2023}, this contamination is suspected to have been introduced into the raw material melt during crystal growth: Gd$_2$O$_3$ powder, the main raw material of GAGG, was measured to contain $<16.30$~mBq/kg (90\% C.L.) of $^{238}$U by HPGe, suggesting that the most likely source of contamination is the zirconia (ZrO$_2$) insulation used in the furnace, which contains approximately 2000~Bq/kg of $^{238}$U. We are therefore currently developing a high-purity insulator available for high-purity GAGG crystal growth for PIKACHU.

As the internal $^{238}$U contamination is reduced, the external backgrounds will become the next dominant contribution in the $\beta/\gamma$-ray energy region, as is already visible in Fig.~\ref{fig:FittingResult}. This can be addressed by replacing the current PMT with a low-background PMT such as a R11065-20A, which has a 64-mm-diameter photosensitive window (Hamamatsu Photonics)~\cite{R11065-20A}. This replacement is expected to provide not only a reduction of the external background but also an improvement in energy resolution, since the matched diameter eliminates the need for a light guide between the GAGG crystal and PMT, thereby increasing the photon collection efficiency. Data taking with this PMT has already begun at the Kamioka underground laboratory. Another option is to use a Silicon Photomultiplier (SiPM) sensor, which contains no glass and therefore eliminates the dominant source of $^{40}$K contamination present in conventional PMTs. Furthermore, the typical photon detection efficiency of a SiPM at the GAGG luminescence wavelength is approximately 40\%, roughly twice that of a conventional PMT, offering a further improvement in energy resolution. However, the intrinsic radioactivity of SiPM devices must be carefully characterized before deployment.

If the improvements described above are achieved in Phase~1 and the projected Phase~2 sensitivity, based on $\mathcal{O}(100)$~kg of crystals and a measurement period of less than ten years, reaches the quasiparticle-random-phase-approximation (QRPA)-predicted half-life of $7.4\times10^{20}$~yr~\cite{Hinohara2022}, the experiment will proceed to Phase~2, which aims to discover $^{160}$Gd $2\nu2\beta$ decay. The feasibility of this goal is supported by a simple sensitivity scaling estimate: with 100~kg of GAGG crystals and a background level one order of magnitude lower than that of Crystal~\#1, the required sensitivity could be reached within approximately two years of data taking.

\section{Conclusion}\label{sec:Conclusion}
We have presented a background characterization for the search for two-neutrino double-beta ($2\nu2\beta$) decay of $^{160}$Gd using Gd$_3$Al$_2$Ga$_3$O$_{12}$ (GAGG) scintillators in the PIKACHU experiment Phase~1. Two GAGG crystals of 2.436~kg and 0.754~kg were operated for 3102~h and 4173~h, respectively, at the Kamioka underground laboratory.

The experimental backgrounds were investigated through a combination of delayed-coincidence (DC) analysis and spectral fitting. The DC analysis identified decay sequences in the uranium and thorium decay chains, providing activity constraints that were subsequently incorporated into the spectral fits as penalty terms. GEANT4-based background models incorporating the measured detector responses --- including the energy resolution, quenching response, and nonproportional scintillation behavior --- were constructed for each crystal. For Crystal~\#4, a dedicated surface contamination model was developed to account for $^{238}$U, $^{234}$U, and $^{226}$Ra deposition on the crystal surface, which was identified through the characteristic spectral distortion of the $^{214}$Po $\alpha$-ray peak. The resulting background models reproduced the observed $\alpha$- and $\beta/\gamma$-ray energy spectra for all datasets with no statistically significant excess attributable to $^{160}$Gd $2\nu2\beta$ decay. As a byproduct of the background characterization, a lower limit on the $^{160}$Gd $2\nu2\beta$ half-life of  $T_{1/2}^{2\nu} \geq 2.32\times10^{19}~\mathrm{yr}\quad\text{at 90\% C.L.}$ was obtained, which is comparable to the best existing limit from a GSO-based search~\cite{Danevich2001}.

The correlation analysis between the signal and background parameters revealed that the $^{40}$K and $^{176}$Lu activities are strongly anti-correlated with the $2\nu2\beta$ signal in the analysis energy window, limiting the achievable sensitivity with the current crystals. Eliminating $^{176}$Lu contamination through the use of a dedicated growth crucible and constraining the $^{40}$K activity via ICP-MS measurement are identified as the two most important steps toward improved sensitivity in Phase~1. Combined with the reduction of the dominant internal $^{238}$U contamination through furnace insulator replacement and the deployment of low-background photosensors, these measures are expected to bring the Phase~1 sensitivity to the level predicted by QRPA nuclear matrix element calculations once crystals of $\mathcal{O}(100)$~kg with the required radiopurity are realized.

\begin{acknowledgments}
We gratefully acknowledge Kamioka Mining and Smelting Co., Ltd. for managing the Kamioka underground laboratory, as well as the laboratory staff for their support in carrying out our experiment. This work was supported by JSPS KAKENHI (Grant Numbers JP22H04570, JP23H01196, JP23K25892, JP25H02168, and JP24H00225), JST SPRING (Grant Number JPMJSP2124), and a JSPS Research Fellowship DC2 (Grant Number JP25KJ0688). Financial support was also provided by SANKO Co., Ltd. The experiment was performed at the Kamioka Extremely Rare-phenomena and NEutrino-research Laboratory (KERNEL).
\end{acknowledgments}

\appendix


\bibliography{apssamp}

\end{document}